\documentclass{aa}

\usepackage{graphicx}
\usepackage{txfonts}

\usepackage[switch]{lineno}

\begin{document}
    \title{The radiative subpulse modulation and spectral features of PSR B1929$+$10 with the whole pulse phase emission}

   \author{Zhengli Wang \inst{1}
          \and
          Jiguang Lu\inst{2,3}
          \and
          Weiyang Wang\inst{4}
          \and
          Shunshun Cao\inst{5}
          \and
          Jinchen Jiang\inst{2}
          \and
          Jiangwei Xu\inst{2}
          \and
          Kejia Lee\inst{5,6}
          \and
          Enwei Liang\inst{1}
          \and
          Hongguang Wang\inst{7}
          \and
          Renxin Xu\inst{5,6}
          }
    \institute{Guangxi Key Laboratory for Relativistic Astrophysics, School of Physical Science and Technology,
              Guangxi University, Nanning 530004, China
         \and
             National Astronomical Observatories,
             Chinese Academy of Sciences, Beijing 100012, China
        \and
            Guizhou Radio Astronomical Observatory, Guiyang 550025, China
        \and
            School of Astronomy and Space Science, University of Chinese Academy of Sciences, Beijing 100049, China
        \and
            Department of Astronomy, School of Physics, Peking University, Beijing 100871, China
        \and
            Kavli Institute for Astronomy and Astrophysics, Peking University, Beijing 100871, China
        \and
            Department of Astronomy, School of Physics and Materials Science, Guangzhou University, Guangzhou 510006, China \\
            \email{lujig@nao.cas.cn, lew@gxu.edu.cn, r.x.xu@pku.edu.cn}
             }
    \authorrunning{Z.L.Wang et al.}

  \abstract
   {The emission mechanism of pulsars is still not well understood. Observations of their intrinsic radio emission and polarization would shed light on the physical processes of the pulsars, such as the acceleration of the charged particles and the radio wave propagation in the pulsar magnetosphere, the location of the radio emission, and the geometry of emission.}
   {To measure the radio emission characteristics and polarization behaviors of the normal and bright pulsar PSR B1929$+$10, we carry out a long-term observation to track this pulsar. Features of its intrinsic emission help us understand the emission mechanism.}
   {In this work, we observe the nearby pulsar, PSR B1929$+$10, using the Five-hundred-meter Aperture Spherical radio Telescope (FAST). This long-term observation includes 110 minutes. A high-precision polarization calibration signal source is required and is carried out in this observation.}
   {We find, for the first time, two new emission components with an extremely weak observed flux density of about $10^{-4}$ of the magnitude of the peak radio emission of PSR B1929$+$10. Our results show that the intrinsic radio emission of PSR B1929$+$10 covers the 360$^{\circ}$ of longitude, demonstrating that this pulsar is a whole 360$^{\circ}$ of longitude emission pulsar. We find at least 15 components of pulse emission in the average pulse profile. Additionally, we identify 5 modes of subpulse modulation in different emission regions, which differ from the pulse components. Moreover, the narrowband emission feature and the frequent jumps in the observed linear polarization position angle (PPA) are also detected in the single pulse of this pulsar. To understand the magnetosphere of this pulsar, we analyze the observed PPA variations across the whole 360$^{\circ}$ of longitude and fit them using the classical rotating vector model (RVM). For the best-fit model, the inclination angle, $\alpha$, and the impact angle, $\beta$, of this pulsar, are 55$^{\circ}$.62 and 53$^{\circ}$.47, respectively. Using the rotating magnetosphere approximation of the magnetic dipole field, we investigate the three-dimensional pulsar magnetosphere and the sparking pattern on the polar cap surface. Our analysis indicates that the extremely narrow zone of the polar cap, which is associated with a high-altitude magnetospheric region, is responsible for the weak emission window. This pulsar has extremely high-altitude magnetospheric radio emissions.}
   {}


   \maketitle

\section{Introduction}
    The normal $\sim$ 226\,ms PSR B1929$+$10 with a distance of about 0.331\,kpc \citep[e.g.,][]{2002ApJ...571..906B,2004MNRAS.353.1311H} has many follow-up observations, these observations aim to study its magnetosphere and emission physics since it was first discovered in 1968 \citep{1968Natur.220..753L}. Previous observation suggests that this pulsar has intrinsic emission across most of its profile \citep[e.g.,][]{1985MNRAS.212..489P,1990ApJ...361L..57P,1997JApA...18...91R,2001ApJ...553..341E,2021ApJ...909..170K,2023RAA....23j4002W}. The unusually wide observed profile (most or all of the pulse period) gives insight into the emission mechanism and challenges the understanding of the acceleration of the charged particles in the magnetosphere. Moreover, PSR B1929$+$10 has thermal emission behavior, and its optical emission behavior with a power-law feature was detected by \cite{2002ApJ...580L.147M}. These emission features would help us understand the e$^{\pm}$ pair formation and the acceleration mechanism close to the emission mechanism of this pulsar. However, understanding the emission physical mechanism and the geometry of the magnetosphere requires the study of the intrinsic radio emission from the profile longitudes with weak emission \citep{1995JApA...16..107M} of this pulsar.
    \par In addition, the feature of polarization emission in most or all of the pulse longitude of PSR B1929$+$10 would provide insight into the study of the pulsar magnetosphere and the understanding of emission physics. Previous polarization observations and the pulsar magnetosphere based on the RVM geometry of this pulsar suggested that the portions of a single cone that is almost aligned with the spin axis are responsible for the radio emission of this pulsar that has an unusually wide average pulse profile \citep[e.g.,][]{1988MNRAS.234..477L,1990ApJ...361L..57P,1991ApJ...370..643B,2001ApJ...553..341E}. The measurement of actual linear polarization emission of the pulse window \citep{1995JApA...16..107M}, particularly in the profile longitude with extremely weak emission is surely required for this pulsar. Moreover, the shape and the size of the polar cap of the pulsars are highly sensitive to the magnetosphere structure under the rotating dipole approximation \citep[e.g.,][]{1995ApJ...438..314R,2000ApJ...537..964C,2004ApJ...606L..49Q,2008ApJ...680.1378H,2019SCPMA..6259505L,2024ApJ...963...65W}. Moreover, the polar cap shape influences the structure of the inner polar gap that is responsible for the creation and the acceleration of the electron-positron e$^{\pm}$ pairs \citep[e.g.,][hereafter RS75]{1975ApJ...196...51R}.
    \par Section \ref{sec1} describes the 110 minutes observation for PSR B1929$+$10 carried out by the FAST and data processing. In Section \ref{sec2}, we present the detection of the radio emission feature of this pulsar. The pulsar's magnetosphere structure is based on the RVM solution for the observed PPA presented in Section \ref{sec3}. We calculate and investigate the emission physics of this pulsar in Section \ref{sec4}. In Section \ref{sec5}, we present general discussions and summarize the main conclusion in Section \ref{sec6}.

\section{Observation and Data Reduction}\label{sec1}
   We carried out the 110-minute polarization observation of PSR B1929$+$10 with FAST on MJD 59904 (November 21, 2022) using the 19-beam receiver system. This receiver system covers the frequency range of 1000 to 1500\,MHz \citep[][]{2019SCPMA..6259502J,2020RAA....20...64J}. The raw data were sampled with a time resolution of 49.152\,$\mu$s and recorded in the 8-bit-sampled search mode PSRFITS format \citep[][]{2004PASA...21..302H}. The frequency channel is 4096, which corresponds to a frequency resolution of about 0.122\,MHz. In this observation, we chose the high level of 10\,K of the noise diode as the polarimetric calibration signal source. To obtain a precise polarization solution and then study the polarization emission feature in the single pulse of this pulsar, the noise diode signals of 30\,s were injected after each observation length for 30 minutes was carried out in our observation. The \textsc{DSPSR} software package \citep[][]{2011PASA...28....1V} was adopted to fold the raw data. The ephemeris of the pulsar was obtained from the ATNF Pulsar Catalogue \citep{2005AJ....129.1993M}. The folded data includes a total period of about 28000, and each period contains the pulse phase bin of 1024. The radio frequency interference (RFI) is mitigated using the dynamic spectrum of two-dimensional frequency-time to eliminate the impact of RFI on the pulsar radio emission. However, due to the event that the reduction in the observed flux density by $2-3$ orders of magnitude in these data reported by \cite{2024ApJ...976L..22W}, we exclude 43 minutes of data and only use the observation length of 67 minutes, which includes the period of about 17000, to analyze and obtain the average pulse profile.

   \begin{figure*}
       \centering{\includegraphics[width = 1.\textwidth]{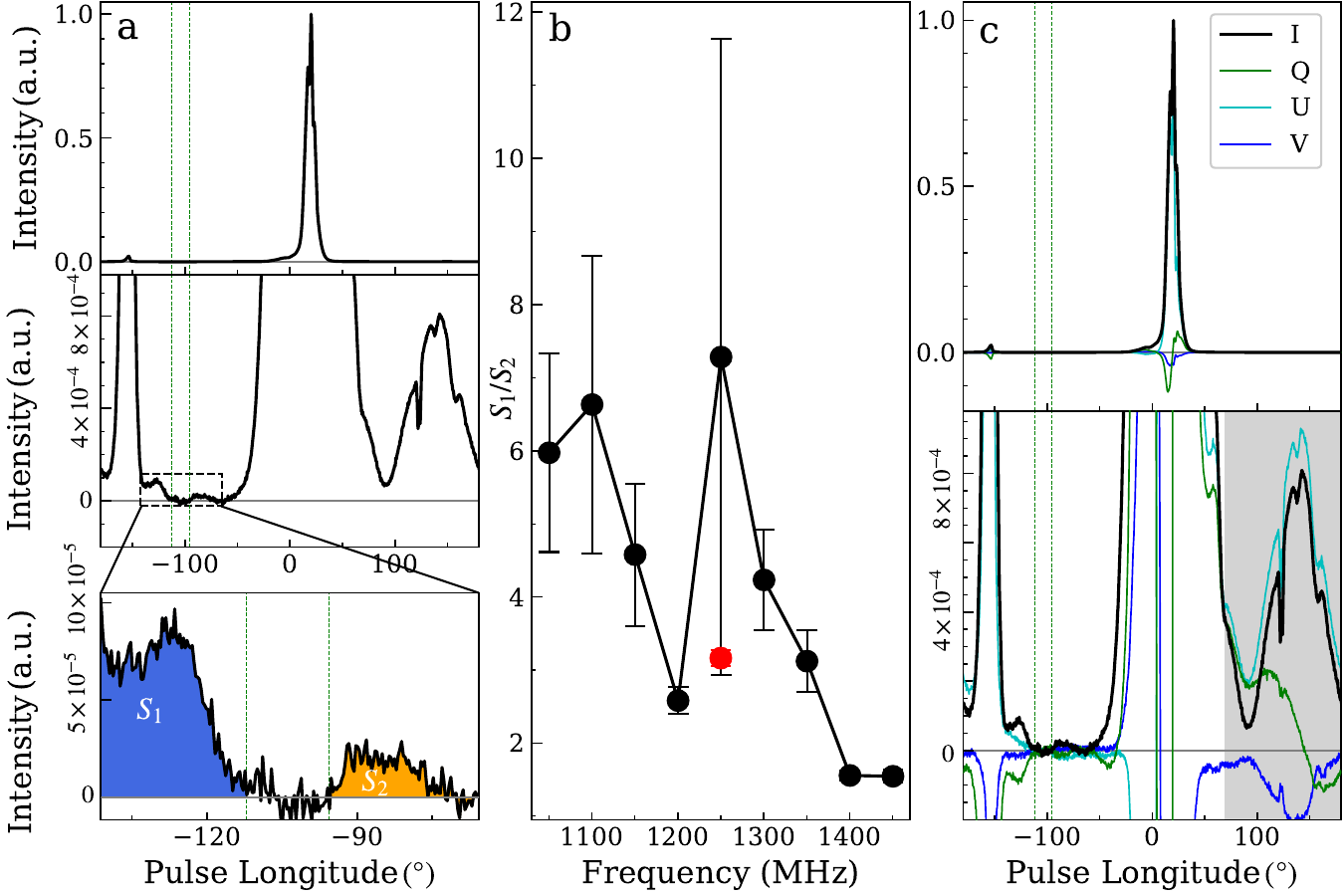}}
       \caption{Radio emission feature of PSR B1929$+$10. (a) The averaged pulse profile is included in the top panel, and the corresponding $\times 1000$ expanded scale view is plotted in the middle panel. To unravel the profile longitudes with extremely weak emission at the pulse longitude range $-150^{\circ}$ to $-70^{\circ}$ even further, a detailed view of the rectangle region in the dashed-dotted box is shown in the bottom panel. The labeled $S_1$ and $S_2$ correspond to the areas of the first and second new pulse components in the left (blue region) and the right (orange region), respectively. The region between the two vertical green dashed lines is believed to be the baseline position, and the solid gray line represents the baseline position. The pulse profile has been normalized by the peak radio emission of the average pulse. (b) To identify whether the two new pulse components are still visible as similar to the result of the bottom panel (a) at nine narrow bands, we calculate the ratio between the $S_1$ and $S_2$, $S_1/S_2$, at nine narrow bands and plot them in this subpanel. The red dot corresponds to the ratio $S_1/S_2$ for the two new pulse components of the average pulse profile. (c) The observed profiles of Stokes parameters for PSR B1929$+$10 based on the conventional baseline subtraction. We plot the total intensity (the Stokes $I$ in the black), the Stokes $Q$ in the green, the Stokes $U$ in the cyan, and the circular polarization intensity (the Stokes $V$ in the blue) in the top panel. To reveal the observed profiles in the weak emission region in more detail, the $\times$1000 expanded scale view is included in the bottom panel. The vertical axis is the same as the vertical axis of panel (a). Same as panel(a), the region between the two vertical green dashed lines is believed to be the baseline position and is then subtracted. The horizontal gray line represents the baseline position. And the vertical gray shadow shows that an unphysical phenomenon is that the Stokes $U$ is greater than the total intensity Stokes $I$, due to the conventional baseline subtraction, which is not suitable for this pulsar. See the main text for further details about the plots.}
       \label{f1}
   \end{figure*}

    \begin{table}
        \caption{Pulse components and their spectral indices. To understand the physical mechanisms of these components, a power-law function is used to fit them, and gives the spectral indices. The first column denotes the components, and their corresponding pulse longitude range is given in the second column. The last column describes the spectral index ($\alpha$).}
        \tabcolsep=3cm
        \renewcommand\arraystretch{1.5}
        $$
        \begin{array}{p{2.5cm}p{3cm}p{2.5cm}}
        \hline \hline
        \noalign{\smallskip}
        Components & Pulse Longitude $(^{\circ})$ & $\alpha$ \\
        \noalign{\smallskip}
        \hline
        \noalign{\smallskip}
            $C_1$ \ \   IP & -174 to -151 & 1.64 $\pm$ 0.31 \\
            $C_2$ & -151 to -137 & 2.82 $\pm$ 0.34 \\
            $C_3$ & -137 to -109 & 5.98 $\pm$ 0.65 \\
            $C_4$ & -102 to -63 & 4.19 $\pm$ 1.36 \\
            $C_5$ \ \  Precursor & -63 to -3 & 0.94 $\pm$ 0.30 \\
            $C_6$ \ \  MP  & -3 to 18.5 & 0.90 $\pm$ 0.30 \\
            $C_7$ \ \  MP & 18.5 to 22 & 1.17 $\pm$ 0.30 \\
            $C_8$ \ \  MP & 22 to 53 & 1.57 $\pm$ 0.31\\
            $C_9$ \ \ Postcursor & 53 to 68 & 2.31 $\pm$ 0.31 \\
            $C_{10}$ & 68 to 89 & 2.38 $\pm$ 0.33 \\
            $C_{11}$ & 89 to 120 & 2.61 $\pm$ 0.33 \\
            $C_{12}$ & 120 to 126 & 3.24 $\pm$ 0.37 \\
            $C_{13}$ & 126 to 138 & 2.52 $\pm$ 0.32 \\
            $C_{14}$ & 138 to 158 & 2.34 $\pm$ 0.32 \\
            $C_{15}$ & 158 to 180 & 2.58 $\pm$ 0.33 \\
            \noalign{\smallskip}
             \hline
        \end{array}
        $$
        \label{t1}
    \end{table}

    \begin{figure}
        \centering
        \includegraphics[width=1.\linewidth]{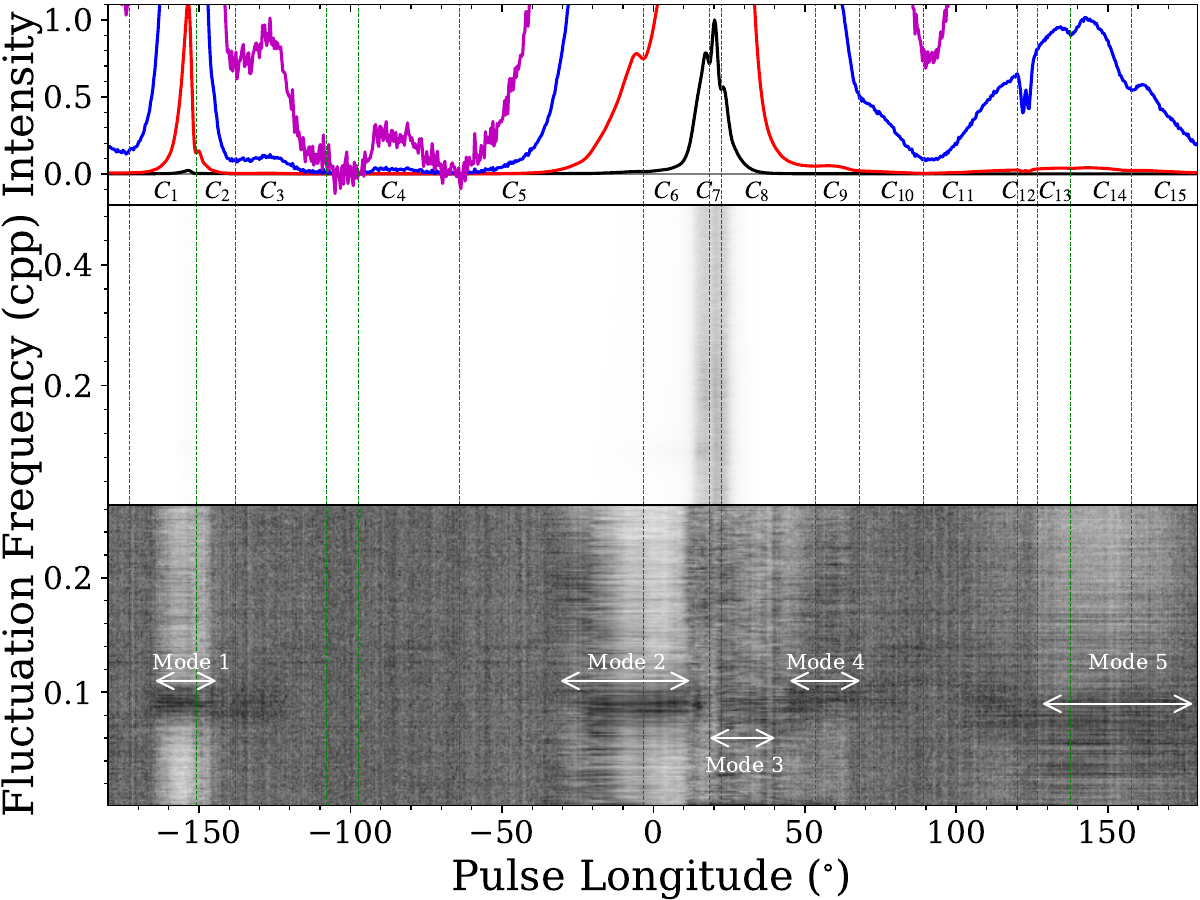}
        \caption{Observed pulse profile and the LRFS for PSR B1929$+$10. The top panel depicts the average pulse profile over about 17000 individual pulses, the intensity is scaled with the peak radio emission. The black curve shows the average pulse profile. To visualize the different pulse emission components, the $50 \times$ scaled zoom view is presented as the red curve, while the blue and magenta curves represent the $1250 \times$ and $10000 \times$ scaled zoom views, respectively. The intensity is measured in units of arbitrary units. The LRFS is calculated and plotted in the middle panel. To show the characteristics of the pulse component in the LRFS and then identify them, we reanalyze the LRFS and utilize the peak value of the LRFS in each pulse longitude to normalize them, respectively. The result is included in the bottom panel, and the arrows indicate the pulse longitude ranges of these modes.}
        \label{LRFS}
    \end{figure}

    \begin{table}
        \caption{The subpulse modulation properties in the different pulse longitudes. The first column depicts the modes. The second and third columns give the corresponding range in the pulse longitude and the fluctuation frequency, respectively. The fluctuation frequency is measured in units of cycles per period (cpp), which corresponds to $P/P_3$. Here, $P$ is the period of the pulsar and $P_3$ denotes the modulation period.}
        \tabcolsep=3cm
        \renewcommand\arraystretch{1.5}
        $$
        \begin{array}{p{1.8cm}p{2.75cm}p{3.5cm}}
        \hline \hline
        \noalign{\smallskip}
        Subpulse Modulation & Pulse Longitude $(^{\circ})$ & Fluctuation Frequency (cpp) \\
        \noalign{\smallskip}
        \hline
        \noalign{\smallskip}
            Mode 1 & -166 to -143 & 0.091 $\pm$ 0.010 \\
            Mode 2 & -32 to 13 & 0.089 $\pm$ 0.012 \\
            Mode 3 & 17 to 41 & 0.085 $\pm$ 0.018 \\
            Mode 4 & 43 to 69 & 0.094 $\pm$ 0.013 \\
            Mode 5 & 122 to 180 & 0.078 $\pm$ 0.024 \\
            \noalign{\smallskip}
             \hline
        \end{array}
        $$
        \label{t2}
    \end{table}

    \begin{figure*}
        \centering
        \includegraphics[width = 1.\textwidth]{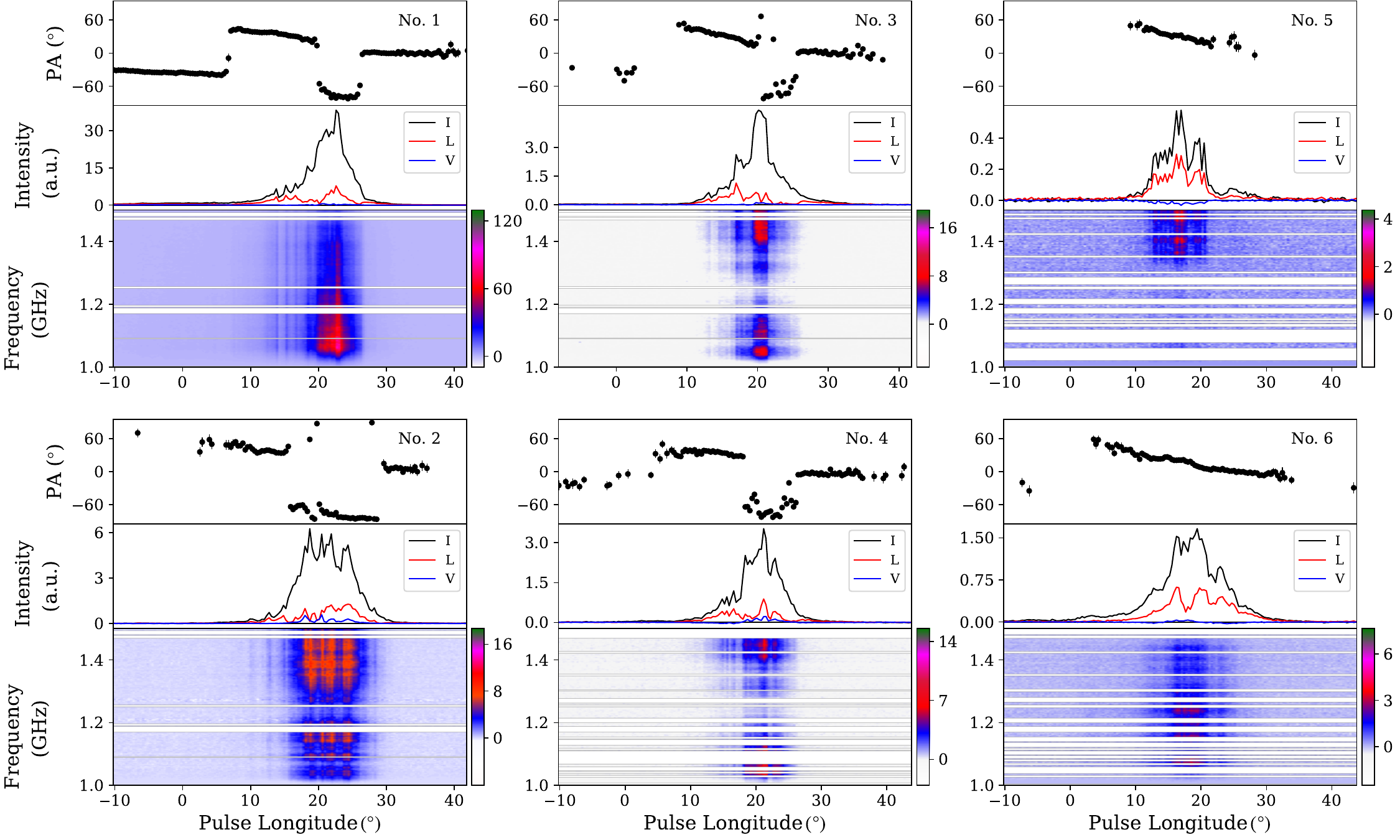}
        \caption{Narrowband emission feature of the selected samples of the single pulses. For each single pulse, the observed PPAs are included in the top panel, and the profile of the total intensity (Stokes $I$ in the black), the linear polarization intensity ($L=\sqrt{Q^2 + U^2}$ in the red), and the circular polarization intensity (Stokes $V$ in the blue) are plotted in the middle panel. To investigate the emission feature of the pulse window in a single pulse further, we plot the dynamic spectra of the flux density as a function of frequency and pulse longitude in the bottom panel. The color bar represents the flux density and is measured in arbitrary units. The horizontal white lines denote the RFI frequency channels and are set to zero. The white indicates the noise floor. We chose the errors in the observed PPAs lower than 5$^{\circ}$ in each single pulse to limit the random error due to the system noise.}
        \label{drift}
    \end{figure*}

    \begin{figure*}
        \centering
        \begin{minipage}[t]{0.48\linewidth}
            \centering{\includegraphics[width = 1.\textwidth]{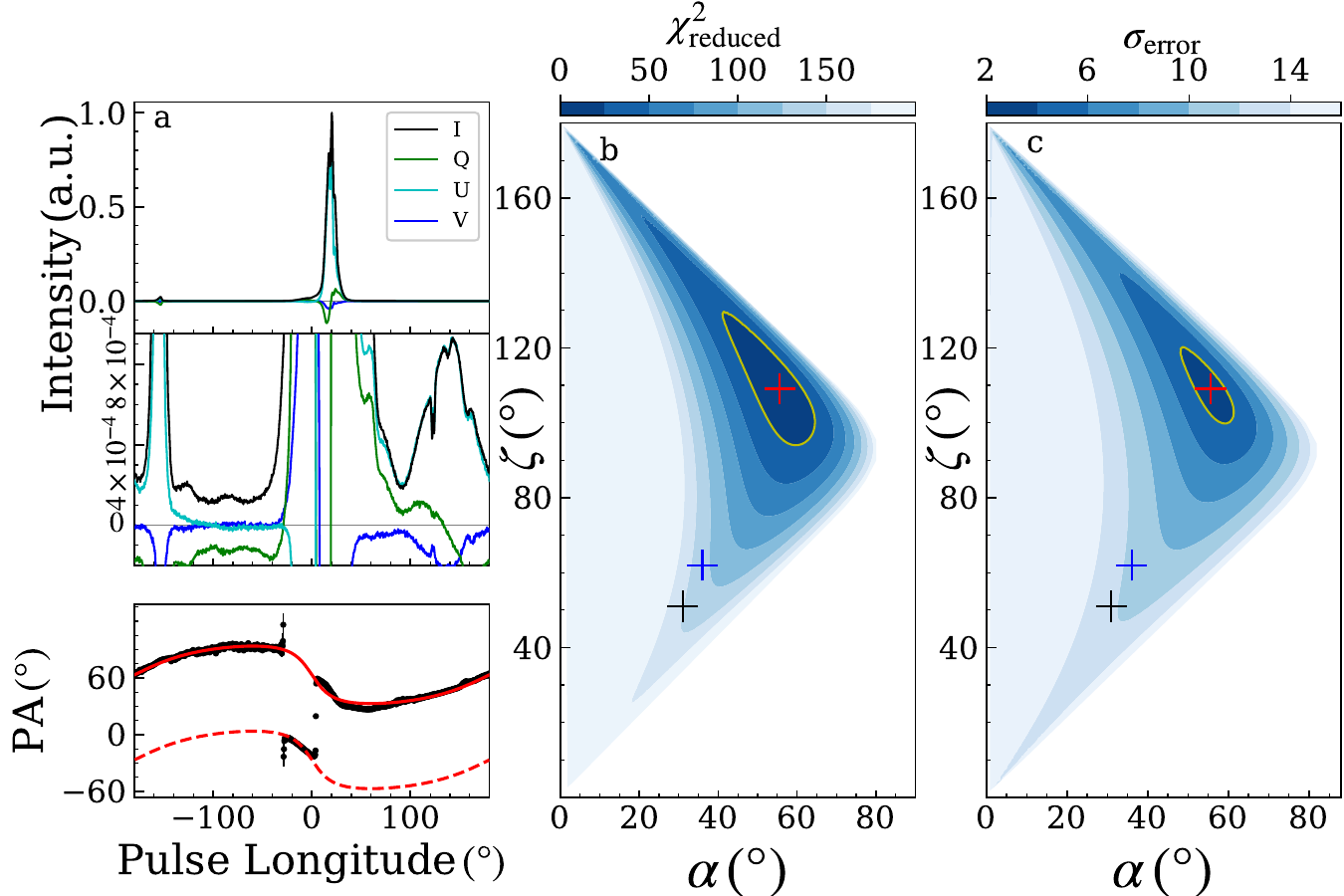}}
            \centerline{(A)}
        \end{minipage}
        \begin{minipage}[t]{0.48\linewidth}
            \centering{\includegraphics[width =1.\textwidth]{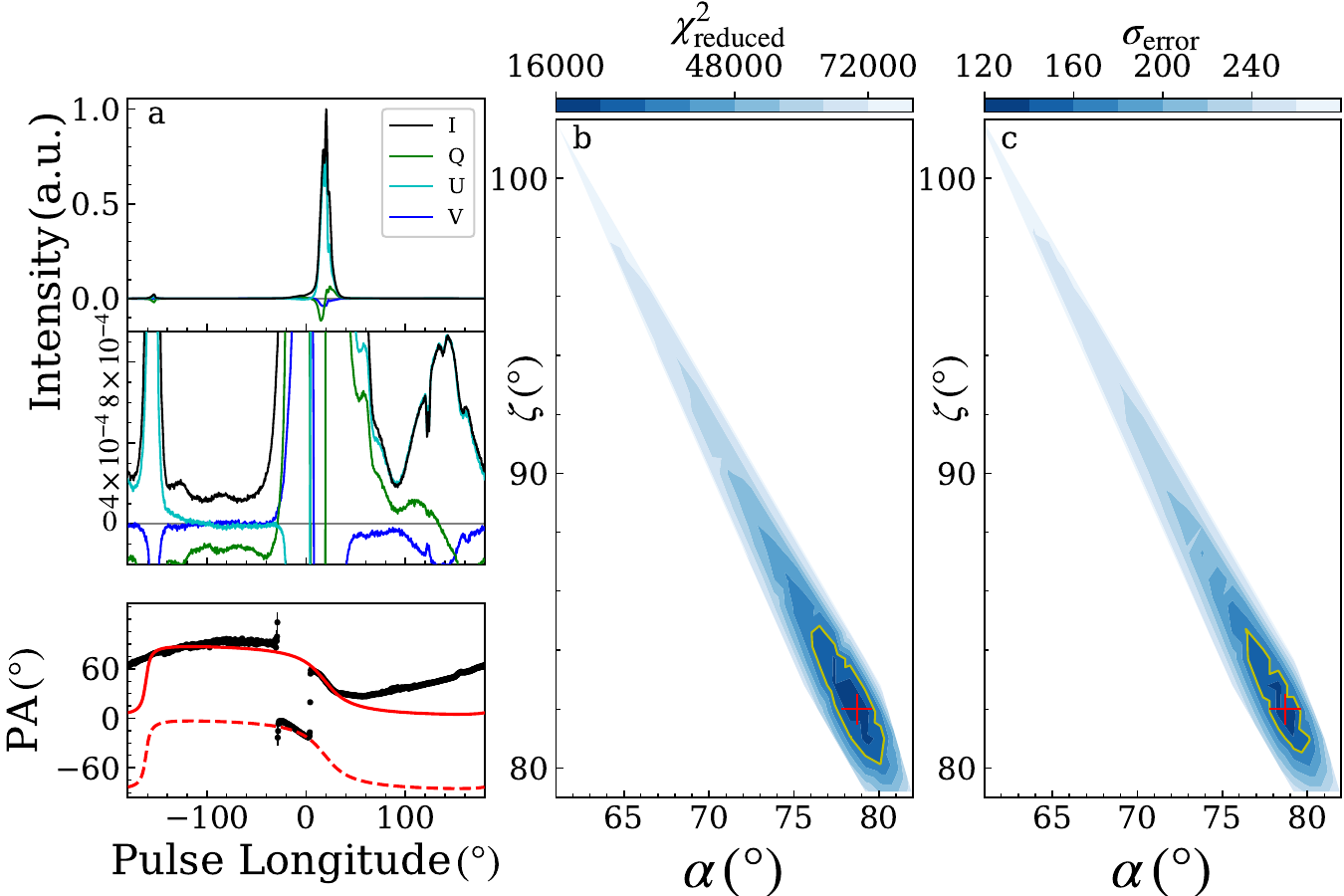}}
            \centerline{(B)}
        \end{minipage}
        \caption{Polarization profiles and the solution of the RVM for the observed PPAs of the average pulse for PSR B1929$+$10. (A) In the left-hand panel, we plot the profiles of the total intensity (the Stokes $I$ in the black), the Stokes $Q$ in the green, the Stokes $U$ in the cyan, and the circular polarization intensity (the Stokes $V$ in the blue) in the top panel. The corresponding $\times 1000$ expanded scale view is included in the middle panel. The horizontal gray line denotes the baseline position in the top and middle panels. The observed PPAs are in black dots and are plotted in the bottom panel. The red curve obtained from the best model fit and its 90$^{\circ}$ offsets are in the dashed red curve. Panel (b) depicts the $\alpha - \zeta$ plane, which shows the $\chi_{\mathrm{reduced}}^2$ from the fitting routine. The $\zeta$ is the viewing angle, which corresponds to the angle between the line of sight (LOS) and the rotation axis. The best RVM solution is included, which is in the red cross, denoting the location of the minimum $\chi_{\mathrm{reduced}}^2$ in the $\alpha-\zeta$ surface. We plot the standard deviation, $\sigma{_\mathrm{error}}$, of the errors between the observed PPAs and the model curve as a function of the $\alpha$ and $\zeta$ in panel (c), which is obtained from the fitting routine. The red cross corresponds to the location of the best RVM solution. For the best RVM model fit, the inclination angle, $\alpha$, and the viewing angle, $\zeta$, of this pulsar are 55$^{\circ}$.62, and 109$^{\circ}$.09, respectively. To compare with previous works of the polarization measurements for this pulsar, we plot the result of $\alpha=36^{\circ}$ and $\zeta=62^{\circ}$ in the blue cross reported by \cite{2001ApJ...553..341E}. While the black cross corresponds to the result of $\alpha=31^{\circ}$ and $\zeta=51^{\circ}$ given by \cite{1990ApJ...361L..57P}. See the main text for further discussion about the polarization measurement. The solid yellow contour line corresponds to the 95 percent confidence regions. (B) Same as panel(A), but the errors of the PPAs are considered when using the RVM to fit the observed PPAs in the full longitude. The best model fit indicates that the value of $\alpha$, the value of $\zeta$, and the steepest gradient of the RVM, $\phi_0$, are 78$^{\circ}$.70, 82$^{\circ}$.0, and $18^{\circ}.40$, respectively. \label{rvm0}}
    \end{figure*}

    \begin{figure}
        \centering{\includegraphics[width=1.\linewidth]{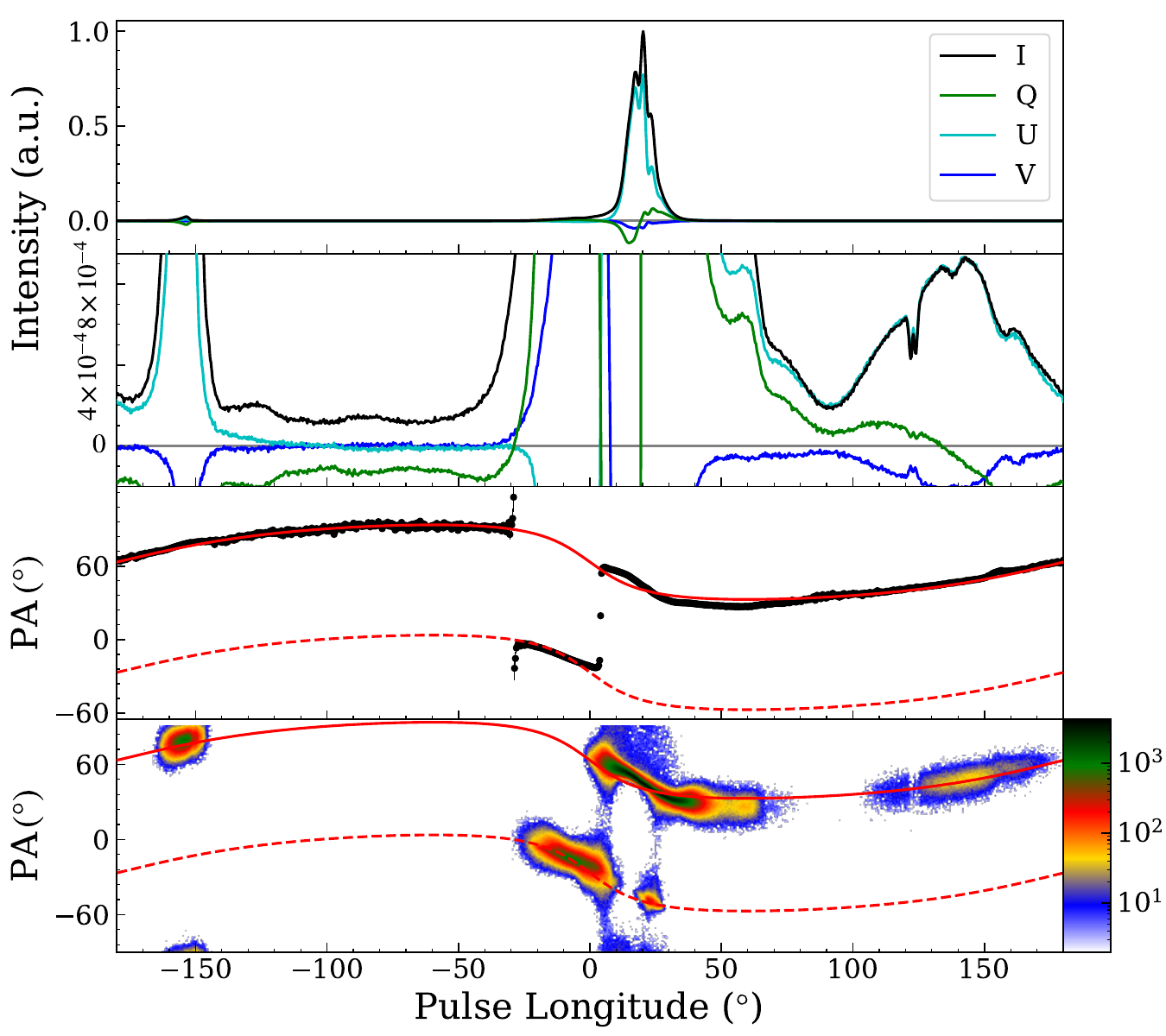}}
        \caption{The polarization emission features for PSR B1929$+$10. We plot the total intensity (Stokes $I$), the Stokes $Q$, the Stokes $U$, and the circular polarization intensity (Stokes $V$) in the top panel. To unravel the features of intrinsic radio emission in the profile longitudes with extremely weak emission and the full longitude, the corresponding $\times 1000$ expanded scale view is included in the second row panel. The profiles of all Stokes parameters have been scaled with the peak intensity of the Stokes $I$. The observed PPAs (black dots) are included in the third-row panel. The best-fitting RVM is in the solid red curve, and its 90$^{\circ}$ offsets are in the dashed red curve. We adjust the steepest gradient of the RVM solution $\phi_0$ to the reference 0$^{\circ}$ of longitude. To reveal the different polarization modes that contribute to the polarization emission feature of this pulsar in different emission regions, we plot the observed PPAs of the single pulse in the bottom panel. The number of individual pulses is indicated by the color bar in the log-scale coordinate view. We have chosen the error bars for the observed PPA that are less than $5^{\circ}$ for each single pulse. Although such a high threshold reduces the counts and selects against the extremely weak emission pulses, this threshold allows us to distinguish the intrinsic radio emission of the distribution from emission introduced by the system noise. The solid and dashed red curves are the same as the third-row panel.}
        \label{rvm_final}
    \end{figure}

    \begin{figure*}
        \centering
        \begin{minipage}[t]{.55\textwidth}
            \centering{\includegraphics[width = 1.\textwidth]{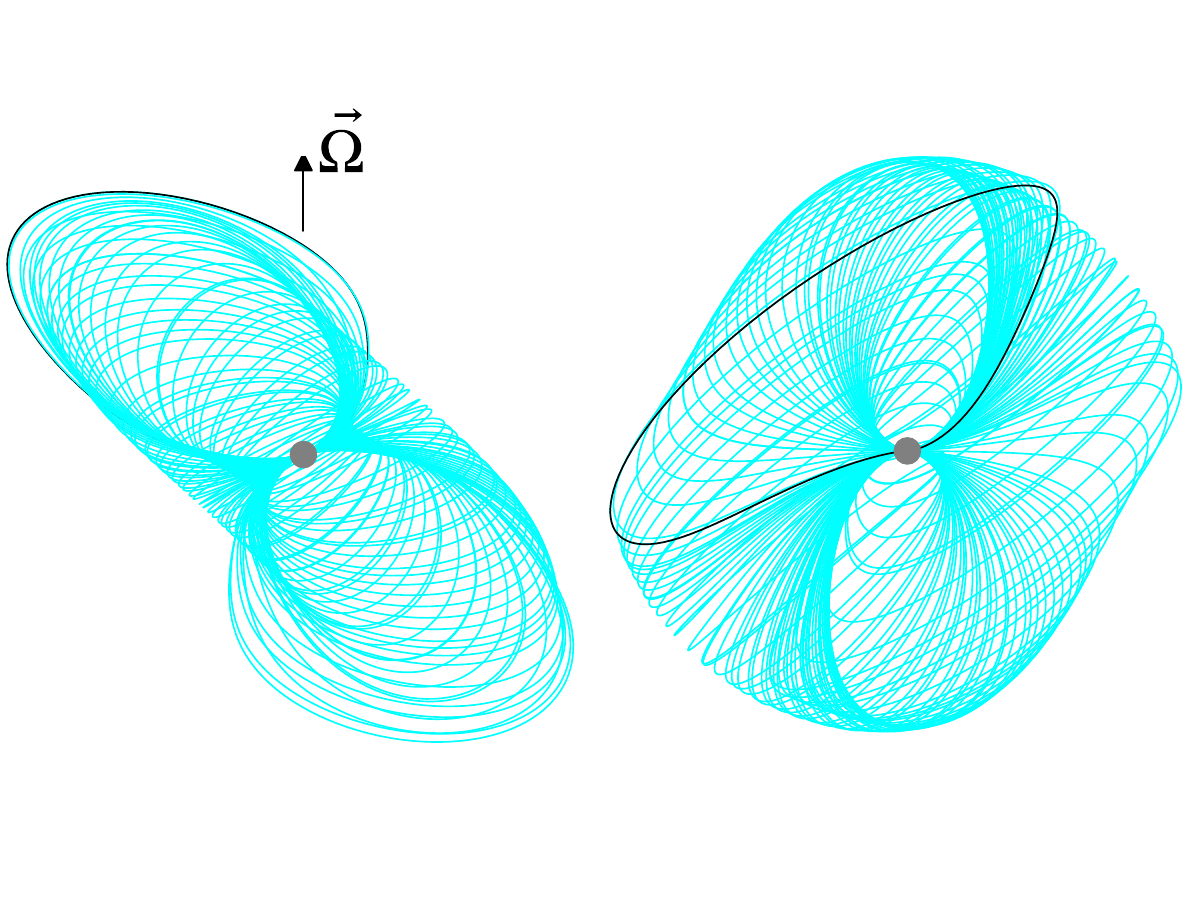}}
            \centerline{(a)}
        \end{minipage}
        \begin{minipage}[t]{.40\textwidth}
            \centering{\includegraphics[width = 1.\textwidth]{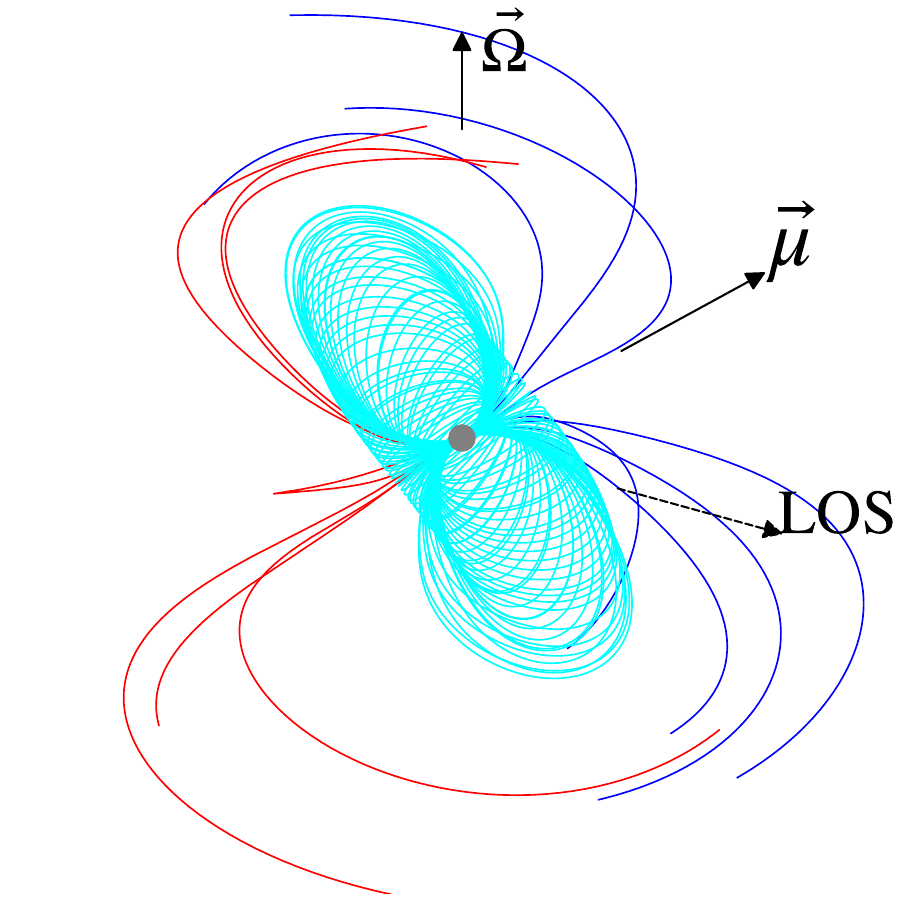}}
            \centerline{(b)}
        \end{minipage}
        \caption{Three-dimensional pulsar magnetosphere for PSR B1929$+$10 assuming the \textsc{RVM} geometry of the inclination angle $\alpha = 55^{\circ}.62$ and the impact angle $\beta = 53^{\circ}.67$ under the rotating magnetosphere approximations. (a) Left: We plot the last closed field lines in cyan, the radius of the pulsar is zoomed to unravel the geometry of the magnetosphere for better clarity. The rotation axis is upward, as labeled $\Vec{\Omega}$. In this view of the plot, the elevation, azimuth, and roll angles are 0$^{\circ}$, -110$^{\circ}$, and 0$^{\circ}$, respectively. (a) Right: A projection looking down in the rotation axis ($x - y$ plane), i.e, the view of the plot corresponds to the elevation, azimuth, and roll angles are 90$^{\circ}$, -90$^{\circ}$, and 0$^{\circ}$, respectively. The magnetic field lines have the phenomenon that they are bunching at low altitude due to the relativistic effect. We also select one of them as the representation and plot it in the black curve. (b) The geometry of the magnetosphere of this pulsar with some open field lines. To better reveal the pulsar magnetosphere, the magnetic field lines that come from the magnetic pole with the inclination angle,$\alpha$, with respect to the rotation axis, are in dark blue curves. Meanwhile, the opposite magnetic poles are in red curves. The magnetic pole and the line of sight are also included, as labeled $\Vec{\mu}$ and LOS. Same as the view in the left panel of (a).}
        \label{pulsar3d}
    \end{figure*}

    \begin{figure}
        \centering{\includegraphics[width=1.\linewidth]{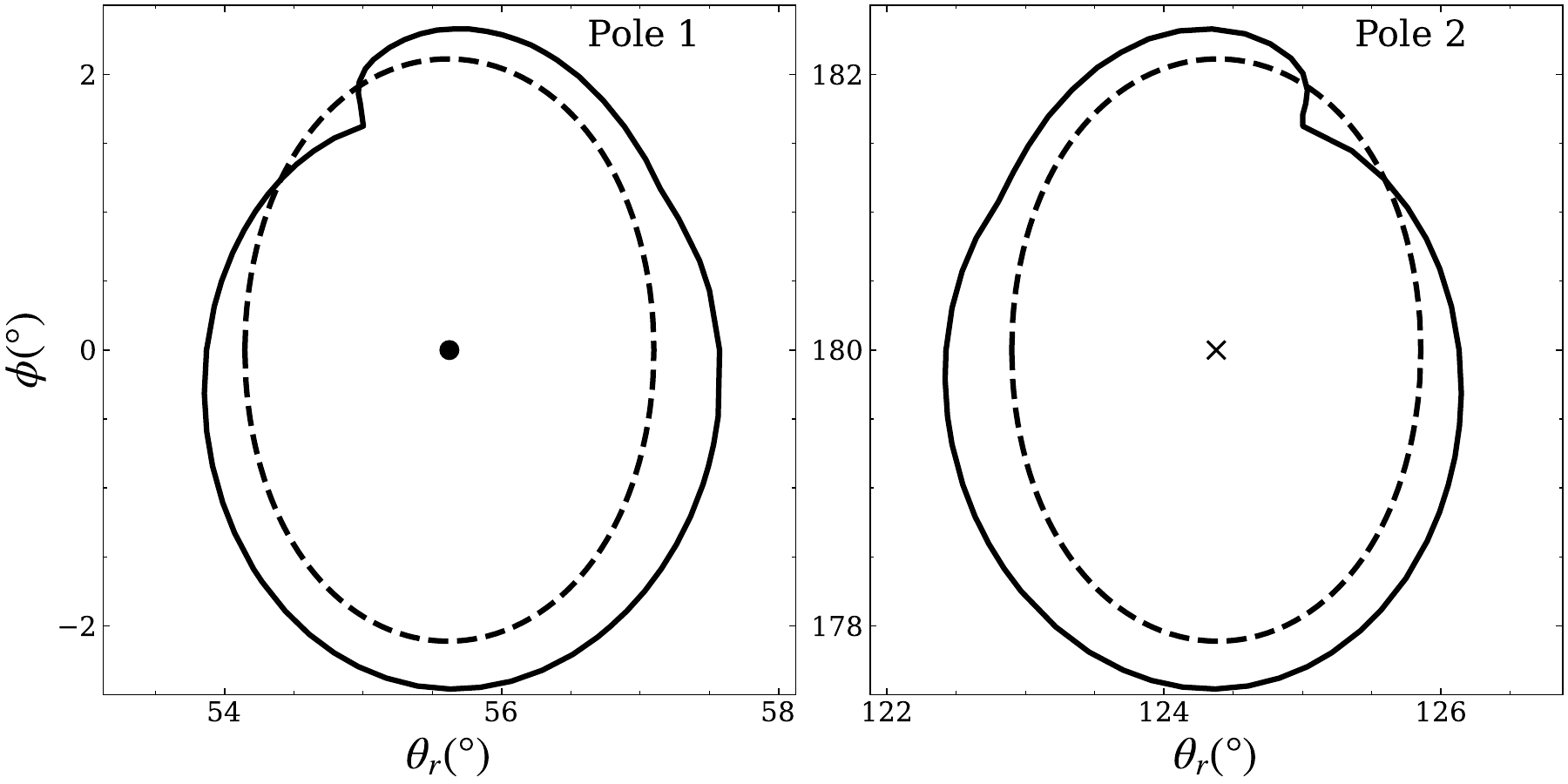}}
        \caption{Polar cap shape for PSR B1929$+$10 assuming the \textsc{RVM} solution of the inclination angle $\alpha = 55^{\circ}.62$ and the impact angle $\beta = 53^{\circ}.47$ under two pole model in rotating magnetosphere approximation. To compare the two magnetospheres in the static and rotating approximations, the polar cap shape of the static dipole is also included, and it is in the dashed black curve. To investigate the emission geometry of this pulsar, the magnetic pole with respect to the rotation axis in the inclination angle $\alpha$ is defined as Pole 1 and labeled with the black dot. Its opposite magnetic pole is defined as Pole 2 and indicated by the black cross. The $\phi$ and $\theta_{\mathrm{r}}$ correspond to the azimuthal and zenith angles with respect to the rotation axis.}
        \label{polarCap}
    \end{figure}

    \begin{figure}
        \centering{\includegraphics[width=1.\linewidth]{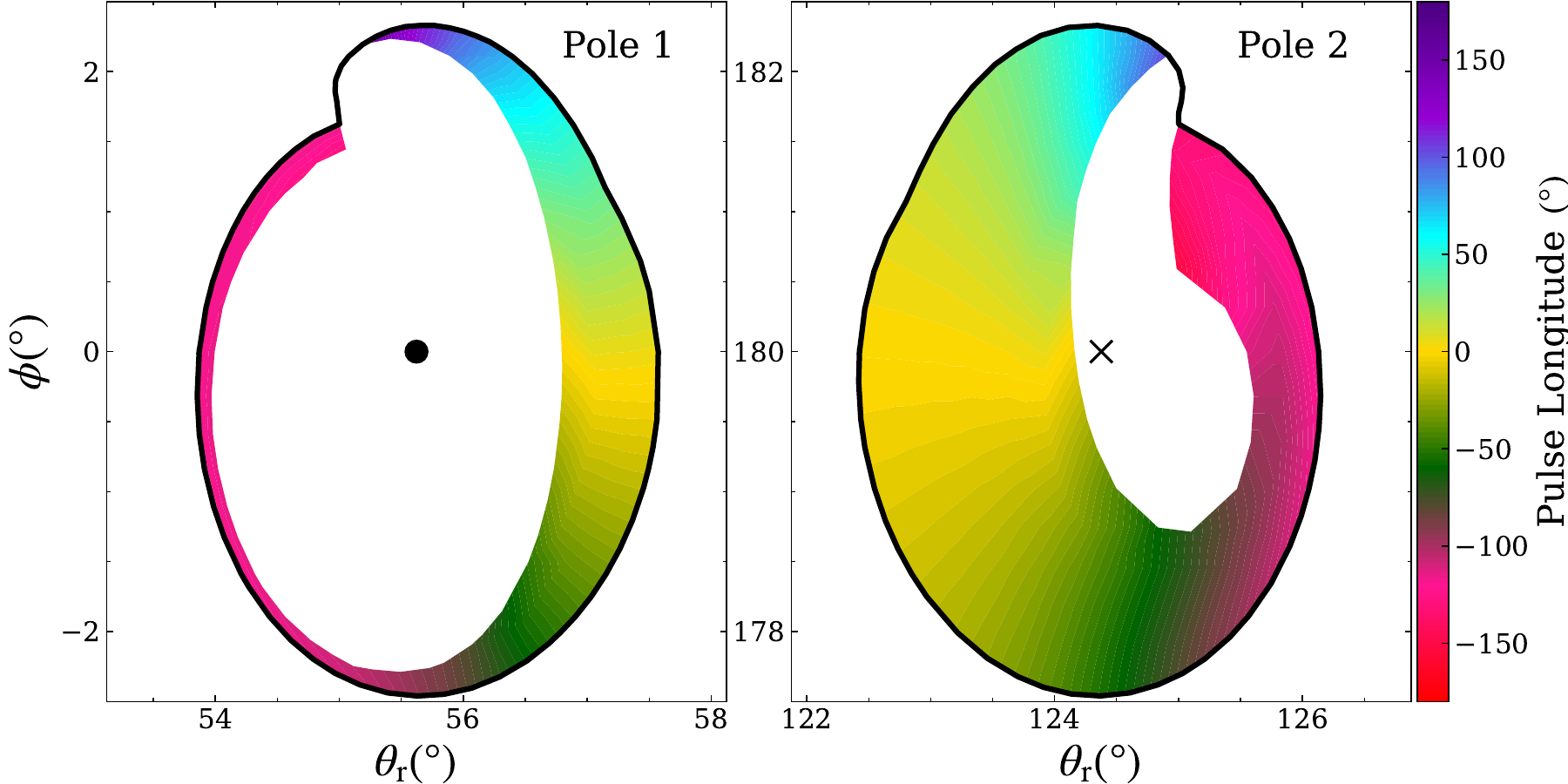}}
        \caption{The sparking pattern on the polar cap surface for PSR B1929$+$10 under the two-pole model. The pulse longitude is marked by color and is responsible for the emission feature. The effect of the aberration and retardation have been taken into account when mapping the pattern of the emission point into the polar cap surface.}
        \label{polarcap}
    \end{figure}

\section{Results}\label{sec2}

\subsection{The whole 360$^{\circ}$ of longitude emission pulsar}
\par The intrinsic radio emission of PSR B1929$+$10 had been detected over extremely wide pulse longitude \citep[e.g.,][] {1985MNRAS.212..489P,1990ApJ...361L..57P,1997JApA...18...91R,2001ApJ...553..341E,2021ApJ...909..170K,2023RAA....23j4002W}. The extremely wide observed profile gives a good opportunity to study the emission physics of this pulsar. Thanks to the high sensitivity of FAST, the intrinsic radio emission signals from two profile longitudes with extremely weak flux density were detected. Fig.~\ref{f1}(a) shows the observed profile of PSR B1929$+$10 over the 67-minute observation, indicating two new emission components are visible through the zoomed view (bottom panel). The flux densities of the new components are extremely weak and are about 10$^{-4}$ of the magnitude of the peak radio emission. The detection of the two emission components is very important to understand the radio emission of this pulsar. It is strong evidence that the radio emission of PSR B1929$+$10 covers the whole 360$^{\circ}$ of longitude, demonstrating this normal pulsar is a whole 360$^{\circ}$ of longitude emission pulsar. The whole 360$^{\circ}$ of the longitude emission feature of this pulsar is different from the full pulse longitude emission pulsar PSR B0950$+$08 \citep{2022MNRAS.517.5560W}.

\par In order to investigate the pulse profiles of the two weak components, we identify whether these two pulse profiles are still visible as similar to that of the average pulse profile (see the bottom panel of Fig.~\ref{f1}(a)) at nine narrow bands. We focus on the shape of the pulse profile, the area of the first pulse component $S_1$ and that of the second pulse component $S_2$, at these bands, and calculate the ratio of $S_1/S_2$. The result is shown in Fig. \ref{f1}(b), showing that these two pulse profiles are still visible with a little fluctuation, similar to that of the average pulse profile. The phenomenon that the trend of the ratio of $S_1/S_2$ is not monotonic at the nine narrow bands is attributed to a large measured error, leading to the variations in the shape of the two pulse profiles at the nine narrow bands. A large measured error is due to the incorrect estimation of baseline position at these bands. This is because the baseline position is directly estimated from the intensity at the region of the two vertical green dashed lines of Fig. \ref{f1} (a). As discussed in the later section, the intensity at the two vertical green dashed lines in the average pulse profile is believed to be the baseline position, yielding an estimated baseline that exceeds the contribution from the ``off-pulse'' region. The red dot corresponds to the ratio of $S_1/S_2$ calculated by the average pulse profile. When the values represented by the black dots are lower than the red dot, it indicates that the second pulse profile becomes clearer in relation to the first pulse component at their corresponding frequencies, compared to the results shown in Fig.~\ref{f1}(a) for the average pulse profile. Conversely, when the values of the black dots are higher than the red dot, it suggests that the first pulse component is more visible compared to the second pulse component.

\par The whole 360$^{\circ}$ of longitude emission feature gives insight into the emission mechanism and challenges the understanding of the phenomenological theories for radio pulsars, particularly in the acceleration of the charged particles in a pulsar magnetosphere and the emission beam geometry. For the whole pulse phase emission pulsar, weak emission windows of the normal pulsar usually come from extremely high altitudes compared to the strong pulse window \citep[e.g.,][]{2022MNRAS.517.5560W,2024ApJ...963...65W} under the magnetic dipole field. The whole 360$^{\circ}$ of the longitude emission feature does not support the model that is based on the inner polar gap of the pulsars proposed by \citet{1975ApJ...196...51R}. This model is used to understand the radio emission feature of the pulsars and is responsible for the primary electrons and the secondary electron-position $e^{\pm}$ pairs production and acceleration in a pulsar magnetosphere. According to the RS75-type models, the inner polar gap is sensitive to the polar cap of the pulsars and exists at a low magnetospheric region with an altitude of about $10^4$\,cm above the polar cap. The RS75-type models can not explain the emission feature of the whole pulse phase rotator with a large inclination angle. Because, with the altitude increases, the strength of the magnetic field and the acceleration zone in the pulsar magnetosphere are a big challenge for the RS75-like pulsar models.

\subsection{Complexity of observed pulse profile and spectral features}
The observed profile of PSR B1929$+$10 is extremely complicated and exhibits a multicomponent emission structure. As shown in Fig. \ref{f1} and the top panel of Fig. \ref{LRFS}, at least $15$ emission components are visible in the average pulse profile. To reveal the fluctuation in flux density of these components with observing frequency and understand their underlying physical processes, we use the power-law function, $I = C f^{-\alpha}$, to fit them and calculate the spectral index $\alpha$. Here, $I$ and $f$ correspond to the observed flux density and the observing frequency, which are measured in units of arbitrary units and GHz, respectively. The results are summarized in Table \ref{t1}. The flux densities of two strong emission windows (i.e., main pulse (MP) and interpulse (IP)) have a flat spectral index compared to that of the weak emission windows. Except for the precursor component of the main pulse, i.e., $C_5$, all visible weak emission components characterize a steep spectral index, and the $\alpha$ is more than 2.30. A more interesting phenomenon is that the two new emission components, i.e., $C_3$ and $C_4$, were first identified by our observation. They have an extremely steep spectral index. The $\alpha$ are up to $5.98$ and $4.19$, respectively. It is worth mentioning that the ``notch-like'' feature is also believed to be a pulse component and corresponds to component $C_{12}$ in Table \ref{t1}.
\par The interstellar scintillation can contribute to the fluctuation in observed flux density. This effect can cause a steep spectral index in the radio emission of the pulsars. For the weak emission windows, the effect of the scintillation may have almost the same contributions. Significantly different spectral indices seen in these emission components may be related to different intrinsic physical processes rather than due to interstellar scintillation. Different spectral indexes for the different emission components in the observed pulse profile may be due to the different physical processes in the pulsar magnetosphere \citep[e.g.,][]{2015ApJ...801L..19P,2022ARA&A..60..495P} such as these emission components come from different magnetospheric regions and polar cap surface. In addition, the measurement of the spectral index for the pulse components may be affected by the effect of the narrowband of the observing frequency.

\subsection{The subpulse modulation properties in the different pulse longitudes}

\par The Longitude Resolved Fluctuation Spectra (LRFS) were proposed to investigate the property of subpulse modulation of the pulsars \citep{1973ApJ...182..245B,2002A&A...393..733E,2006A&A...445..243W}. For the extremely complex profile of PSR B1929$+$10, the subpulse modulation properties in the different pulse components provide insight into the underlying sparking mechanism. We separate the pulse components and investigate their subpulse modulation properties. The results are presented in Fig. \ref{LRFS}, which indicates that various pulse components are associated with different properties. The middle panel shows the LRFS of the whole pulse longitude. To investigate the property of the subpulse modulation in the different pulse longitudes, we reanalyze the LRFS shown in the middle panel of Fig. \ref{LRFS}. In order to avoid the issue of weak radio emissions at certain pulse longitudes, which suffer from a low signal-to-noise ratio, we average ten spectra into a single spectrum along the vertical axis of the LRFS, and then use the peak value of the LRFS at each pulse longitude to normalize the LRFS at that longitude. This normalization yields the power for each pulse longitude that includes at least one peak power point, which does not impact the LRFS of pulse longitudes that exhibit the modulation feature. This is because the modulation feature identifies a distinct region that shows excess power. To unravel the modulation characteristics more thoroughly across the entire pulse longitude, we only focus on a specific portion of the LRFS where the fluctuation frequency is below 0.265, rather than considering the entire LRFS, and then normalize the portion of the LRFS by its peak value at each pulse longitude. The results are shown in the bottom panel of Fig. \ref{LRFS}, where the maximum fluctuation frequency is approximately 0.265. This normalization results in uniform power (i.e., shade of gray) observed at $-100^{\circ}$ and $-50^{\circ}$, where the modulation features are not detected. As shown in the bottom panel of Fig. \ref{LRFS}, the 5 modes of subpulse modulation features are visible, as indicated by the labels. The detection of Modes 1 and 3 agrees with the previous work of this pulsar observed with FAST \citep[e.g.,][]{2021ApJ...909..170K}.

\par From the top and bottom panels of Fig. \ref{LRFS}, the pulse component identified by the eye in the observed pulse profile is not the same as the identification way by the subpulse modulation modes. At least 15 pulse components are visible in the average pulse profile (top panel). However, only 5 subpulse modulation modes are seen in the LRFS (bottom panel). To investigate these modulation modes, we quantify their modulation features and summarize them in Table \ref{t2}. The interpulse and pre/post-cursors of the main pulse display similar characteristics in subpulse modulation, referred to as Modes 1, 2, and 4, which are characterized by a narrow range in the vertical direction of LRFS and have comparable values of the fluctuation frequency. This phenomenon indicates that these modes are stable features with fixed periodicity of modulation in the different ranges of their corresponding pulse longitudes. However, the diffuse modulation phenomenon that corresponds to a wide range in the vertical direction of the LRFS, referred to as Modes 3 and 5, is seen in the main pulse and within the pulse longitude range of 122 to 180$^{\circ}$, respectively. This implies that Modes 3 and 5 have unstable modulation characteristics in the periodicity. Also, the fluctuation frequencies of Modes 3 and 5 are lower than those of Modes 1, 2, and 4. This suggests that different physical processes contribute to the emission features of these pulse longitudes.

\subsection{Narrowband emission feature and frequent jump in observed PPAs in the single pulse}

\par Precise polarization observation provides a good opportunity to investigate the polarization emission feature in the single pulse of this pulsar. We detect the narrowband emission feature, commonly seen in fast radio bursts (FRBs), in some rotation periods of this pulsar. As Fig. \ref{drift} shows, these periods behave as a narrowband emission feature. The period of No. 5 has detectable observed flux density at high frequencies and remains undetectable at low frequencies. For the periods of Nos. 1 and 6, the radio signal was detected at low frequency. This emission feature is usually seen in the single pulse of the pulsars. Another interesting radio emission state is seen in the periods of Nos. 2, 3, and 4, showing that radio emission is a nulling state at the middle frequency. However, the top and bottom band frequencies have a visible emission signal. Also, frequent jumps in the observed PPAs are seen in some periods. The polarization emission for the periods of Nos. 1 and 2 has three jumps in their observed PPAs. The complex polarization phenomenon that behaves as frequent jumps in the observed PPAs in a single pulse suggests that its physical process must be extremely complex. The origin of this phenomenon corresponds to the complex processes of the radio wave propagation in a pulsar magnetosphere \citep[e.g.,][]{1986ApJ...302..120A,1993ppm..book.....B,2001A&A...378..883P,2010MNRAS.403..569W}.
\par The FRBs \citep[e.g.][]{2020Natur.586..693L,2022RAA....22l4003J,2024NSRev..12E.293J} behave as the narrowband emission feature. Recently, a sudden jump in the observed PPAs has been detected in the FRBs \citep[e.g.,][]{2024ApJ...972L..20N}. Similar emission features are detected in the FRBs and the single pulse of this pulsar, implying the physical origin of the FRBs, at least some of them, coming from a complex magnetosphere that is similar to the pulsar magnetosphere. It further indicates that the trigger mechanism of some FRBs is analogous to radio pulsars. However, the huge luminosity difference between these two objects is still a big challenge.

     \begin{figure*}
        \centering
        \begin{minipage}[t]{0.49\linewidth}
            \centering{\includegraphics[width = 1.\textwidth]{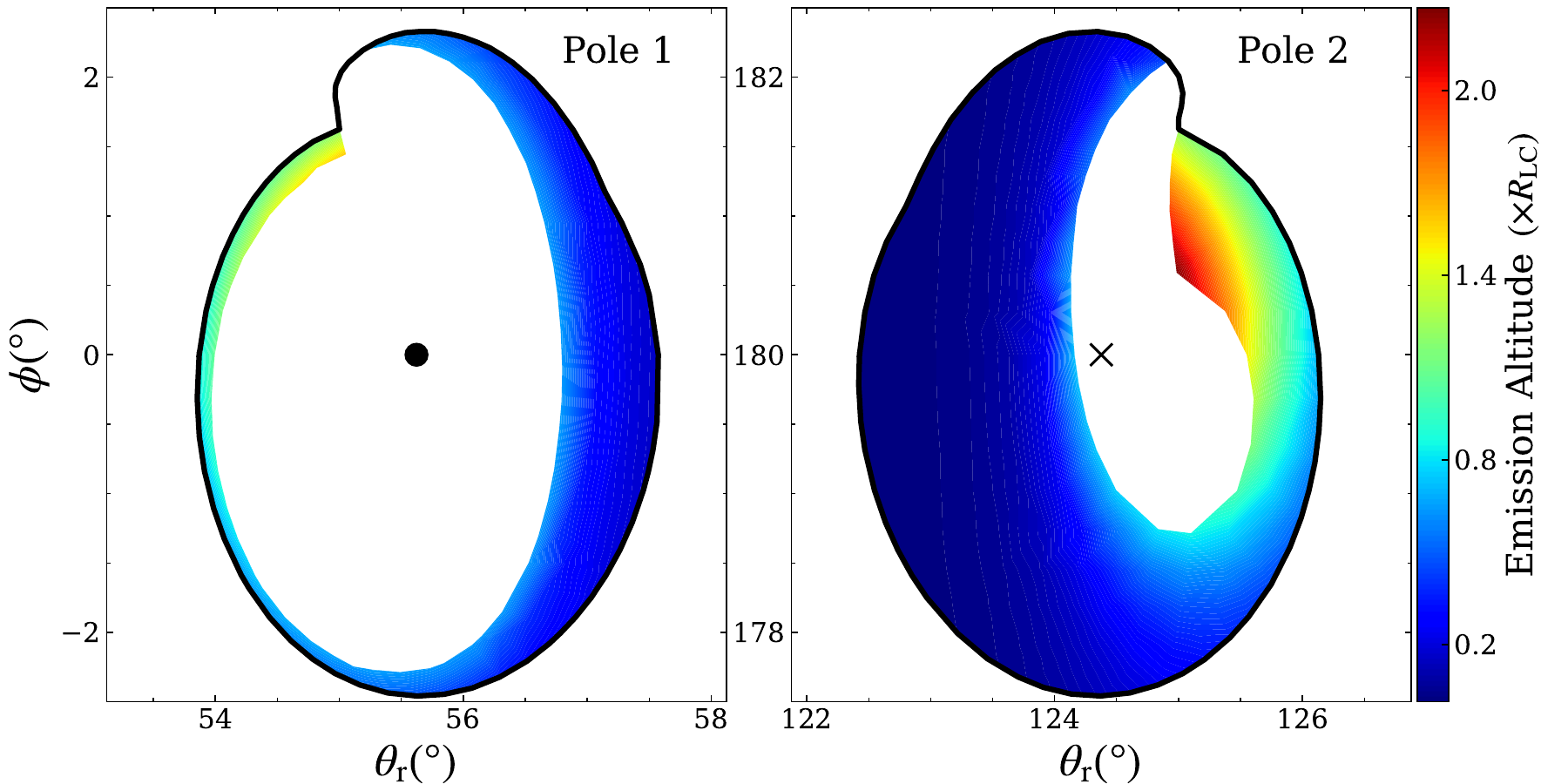}}
            \centerline{(a)}
        \end{minipage}
        \begin{minipage}[t]{0.49\linewidth}
            \centering{\includegraphics[width =1.\textwidth]{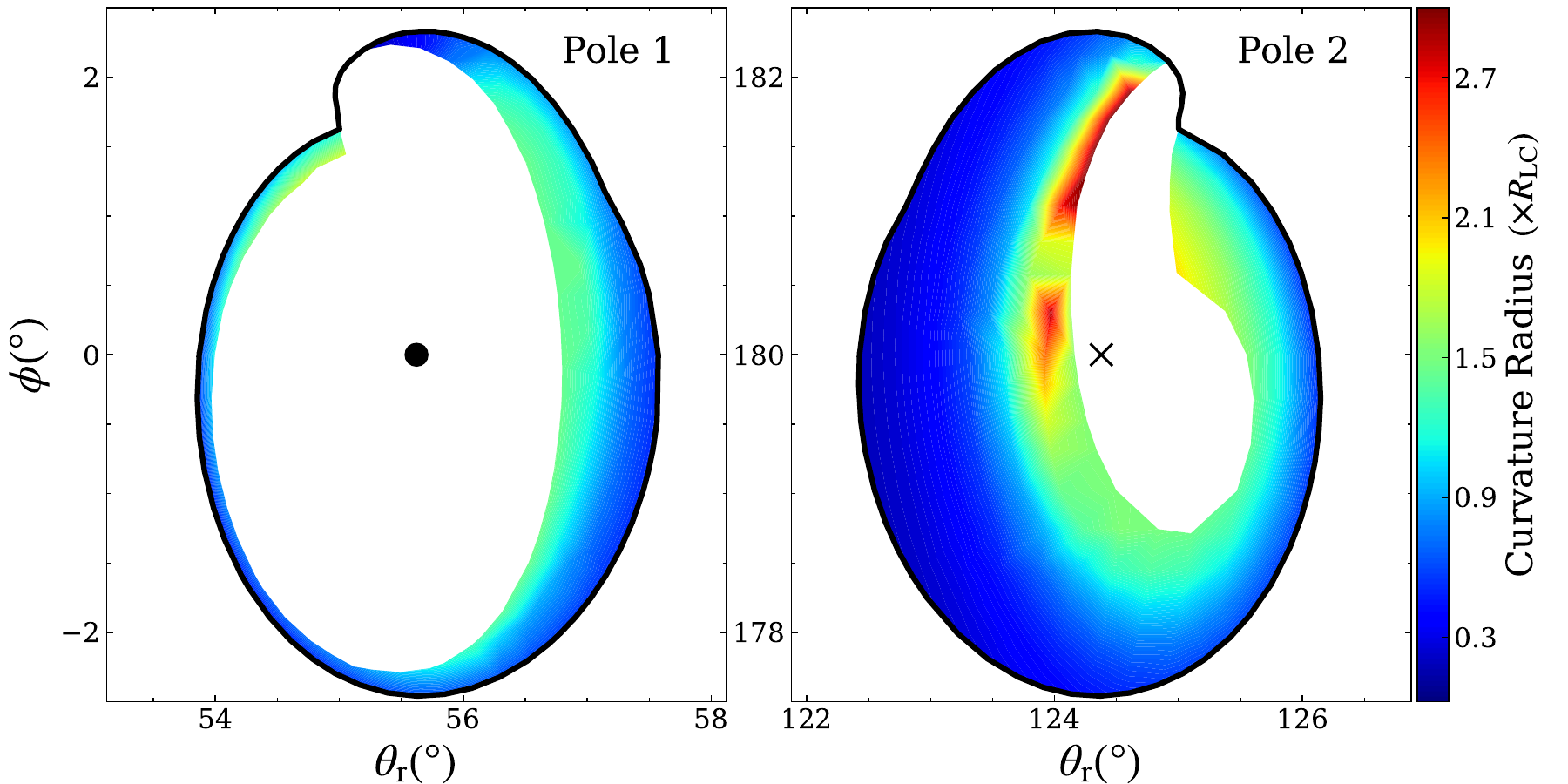}}
            \centerline{(b)}
        \end{minipage}
        \caption{(a)The emission altitude of the emission point for PSR B1929$+$10 under the assumption of the rotating magnetosphere approximation. The aberration effect has been taken into account. The emission altitudes are described by the colors and are measured in units of the light cylinder, $R_{\mathrm{LC}}$, of this pulsar. (b)To further investigate the emission geometry, we calculate and plot the curvature radius of the emission point, as the color bar indicates. The $R_{\mathrm{LC}}$ is about 10807\;km. \label{points}}
    \end{figure*}

\section{The Geometry of the Magnetosphere} \label{sec3}

\subsection{RVM solution and its geometry}
For the whole pulse phase emission pulsar, the determination of baseline position is a big challenge. The conventional baseline subtraction subtracts the intensity of the pulse longitude whose intensity is comparable with the system noise and is far away from the strong emission window (i.e., the main pulse). In this work, we also utilize this method to obtain the observed profiles of this pulsar, the result is shown in Fig. \ref{f1}c. One can see an unphysical phenomenon is that the polarization profiles (Stokes $U$) are higher than the average pulse profile (Stokes $I$) in the pulse longitude range of 70 to 180$^{\circ}$, as the vertical gray shadow indication (bottom). This problem is due to the conventional baseline subtraction is not suitable for this pulsar and subtracts the intrinsic radio signal of the profile longitude with the extremely weak emission. The incorrect baseline subtraction not only affects the observed profiles in the weak emission window but also influences the swing of the observed PPAs in full longitude. \cite{2024ApJ...963...65W} considered the stability of the average pulse profile over long timescale and the polar cap's electric field distribution property, they proposed two tentative estimations of the baseline position for the whole 360$^{\circ}$ of longitude emission pulsar PSR B0950$+$08 \citep{2022MNRAS.517.5560W}. Following Equation (2) in \cite{2024ApJ...963...65W}, we estimate the baseline position of this pulsar and subtract it. The results are shown in Fig. \ref{rvm0}.

\par In Fig. \ref{rvm0}, one can see that highly linear polarized emission is detected over the whole pulse phase of this pulsar. Most of the pulse longitudes where the linear polarized emission characterizes near $100 \%$  polarization factor. This polarization emission property is highly interesting, seen in this whole pulse phase emission pulsar, and hints at the connection between this pulsar and the young pulsars, the magnetars. The polarization emission of the young pulsars and the magnetars usually characterizes highly polarized emission characteristics \citep[e.g.,][]{2006MNRAS.368.1856J,2007MNRAS.377..107K,2010ApJ...721L..33L}. In addition, the polarization emission behavior of this pulsar in the main pulse is significantly different from that of the whole pulse phase emission pulsar, PSR B0950$+$08. PSR B0950$+$08 exhibits a de-polarization phenomenon in the main pulse and characterizes a low linear polarization factor ($\leq 10 \%$) \citep{2024ApJ...963...65W}. This difference in the polarization emission between these two objects is highly interesting since they have similar properties in the observed profile, the distance, the characteristic age, and the powered energy. Detailed discussions will be presented in the later section (Section \ref{sec5}).

\par To investigate the magnetosphere geometry of this pulsar, we fit the observed PPA variations by the RVM \citep{1969ApL.....3..225R} in the full longitude. With the polarization emission signals of the extremely weak emission window detected, the actual deviation from the observed PPA variations and the RVM curve is huge \citep{2004ApJ...609..354M}. The difference results in a large deviation from 1 in the value of $\chi_{\mathrm{reduced}}^2$ and is due to the framework of the geometry based on the RVM is not a good model to predict the PPAs of these pulsars that have a complicated observed profile. The RVM geometry fails to describe the observed PPAs of the pulsars that character the radio signal over most and all of the pulse period in the whole longitude (PSRs B0950$+$08, and J2145$+$0750) \citep{2004ApJ...609..354M,2022MNRAS.517.5560W}. For the whole pulse phase emission pulsar, PSR B0950$+$08, only observed PPAs in the pulse longitude that characterizes a high linear polarization factor ($\geq 30 \%$) can be described by the RVM geometry well \citep{2024ApJ...963...65W}. However, the deviation between the observed PPAs of the whole pulse phase emission pulsar PSR B1929$+$10 and the RVM curve is weak and has a slight value of $\chi_{\mathrm{reduced}}^2$ (about 10) from 1. For all of the pulse period emission pulsars, we use the standard deviation of the errors between the observed PPAs and the model curve, $\sigma_{\mathrm{error}}$, to constrain the RVM solutions and find the best one. We find the $\sigma_{\mathrm{error}}$ is a better one to show the fitting routine (see Fig. \ref{rvm0}A(c)).

\par The results of the RVM solution are shown in Fig. \ref{rvm0}. Compared with the unweighted fit results that are shown in Fig. \ref{rvm0}A, Fig. \ref{rvm0}B shows the results from the RVM solution, which considers the errors of the observed PPAs when fitting the variations in the observed PPAs by the model. This RVM solution suggests a small impact angle, $\beta = \zeta - \alpha$, is about 3$^{\circ}$.3, indicating the LOS of this pulsar is almost aligned with its magnetic axis and corresponds to an extremely small emission beam. However, the huge difference between the observed flux density of the peak radio emission of the main pulse and the interpulse can not be interpreted if the emission geometry of this pulsar, based on the result of Fig. \ref{rvm0}B, is adopted. The observed flux density of the peak radio emission at the main pulse is about $10^3$ times as large as that of the interpulse. This RVM geometry fails due to the different weighted values in the observed PPAs contribution to the geometry of this pulsar at the different emission pulse phases, directly estimating from the errors of the observed PPAs.

\par For the pulsars with the extremely wide observed profile that occupies most or all of the pulse period, the contributions of the observed PPAs in the different pulse longitudes to their magnetosphere geometry are not consistent with the estimation based only on the errors in the observed PPAs in their corresponding pulse longitudes. This estimation yields the observed PPAs in the pulse longitude with small measured errors, making a dominant contribution to the magnetosphere geometry compared to the emission phase, with large errors in the observed PPAs when fitting them by the RVM. As Fig. \ref{rvm0}B(a) shows, only the observed PPAs at the main pulse and interpulse can be described by the RVM curve well. The swing of the observed PPAs at other pulse phases shows a large deviation from the RVM curve. This is because only the observed PPA variations in the two strong emission windows (i.e., main pulse and interpulse) have a dominant contribution when fitting the observed PPAs of this pulsar in the full longitude based on the RVM. Small measured errors in the observed PPAs of these two emission components cause their corresponding observed PPA swings to have huge weighted values compared to other pulse longitudes.

\par For the whole pulse phase emission pulsar, the correct weighted estimation is a big challenge since it may be related to the altitude and curvature radius of the emission location. However, the estimation based on the measured errors in the observed PPAs is still suitable and does work well for the pulsars with the narrow observed profiles (i.e., the duty cycle is about $ 10 \%$). In addition, the $\alpha$ and $\zeta$ have a large covariance when taking the errors in the observed PPAs into account. For the above reasons, we adopt the RVM solution of the inclination angle, $\alpha = 55^{\circ}.62$, and the viewing angle, $\zeta = 109^{\circ}.09$, of this pulsar, to investigate its emission geometry and physics in the later analysis. For the best RVM solution, PSR B1929$+$10 is a pulsar that characters a large inclination angle and is not a nearly aligned geometry \citep[e.g.,][]{1988MNRAS.234..477L,1990ApJ...361L..57P,1991ApJ...370..643B,2001ApJ...553..341E}.

\par A large number of previous works also reported the best RVM solutions for their polarization measurements of this pulsar \citep[e.g.,][]{1988MNRAS.234..477L,1990ApJ...361L..57P,1997JApA...18...91R,2001ApJ...553..341E,2021ApJ...909..170K,2023RAA....23j4002W}. For the best RVM solution of our polarization measurements, we found $\alpha = 55^{\circ}.62$ and $\zeta = 109^{\circ}.09$ for this pulsar. It is important to analyze these results in comparison to previous investigations. Specifically, the best RVM solution reported by \cite{1990ApJ...361L..57P} indicates $\alpha = 31^{\circ}$ and $\zeta = 51^{\circ}$, while \cite{2001ApJ...553..341E} presents values of $\alpha = 36^{\circ}$ and $\zeta = 62^{\circ}$. We plot the results of \cite{1990ApJ...361L..57P} and \cite{2001ApJ...553..341E} in the fitting routine (Fig.\ref{rvm0} A). One can see that the best RVM solution for the observed PPAs swing obtained by this observation is different from previous works. This is because this observation detects the linear polarization emission signal of this pulsar over the whole 360$^{\circ}$ of longitude and uses an unweighted fit.

\par The best RVM solution for this pulsar is shown in Fig. \ref{rvm_final}, the steepest gradient of the RVM solution $\phi_0$ has been adjusted to the reference 0$^{\circ}$ of longitude. One can see that the observed PPA of this pulsar aligns well with the prediction curve of the RVM in most pulse longitudes, as most emission regions exhibit highly linear polarization features. In the bottom panel of Fig. \ref{rvm_final}, an intriguing phenomenon is observed in the polarization emission state of the single pulse. A new 90$^{\circ}$ jump in the observed PPA is detected within the pulse longitude range of 20 to 30$^{\circ}$, as illustrated by the patches, which corresponds to the pulse component $C_8$. This phenomenon is that both 90$^{\circ}$ jumps are seen in the observed PPA of the pre/post cursors of the main pulse may account for the slight deviation observed between the PPA and the RVM prediction curve in the main pulse. This is because the $90^{\circ}$ of the RVM aligns closely with the observed PPA of the main pulse. Compared to the average pulse profile, the observed PPA of individual pulses can provide a clearer reveal of the polarization emission state, particularly in regions of strong emission. However, the weak emission regions suffer from a low signal-to-noise ratio, resulting in a lack of identifiable polarization emission states.

\subsection{Three-dimensional pulsar magnetosphere}

\par The whole 360$^{\circ}$ of longitude emission pulsar is a good opportunity to test the pulsar phenomenological theories. To understand the emission geometry of PSR B1929$+$10 that characterizes the whole pulse phase emission characteristics, we investigate the three-dimensional pulsar magnetosphere of this pulsar. The magnetosphere geometry is based on the assumption of the RVM solution of the $\alpha = 55^{\circ}.62$ and $\zeta = 109^{\circ}.09$ under the rotating magnetosphere approximation of the magnetic dipole field. The three-dimensional rotating magnetosphere model has been investigated in previous works \citep[e.g.,][]{1995ApJ...438..314R,2000ApJ...537..964C,2019SCPMA..6259505L}. Following the vacuum magnetic field $\mathbf{B}$ of the dipole rotator \citep[e.g.,][]{1975clel.book.....J,2000ApJ...537..964C}, we have
\begin{equation}
    \mathbf{B} = - \left[ \frac{\mathbf{\mu}(t)}{r^3} + \frac{\dot{\mathbf{\mu}}(t)}{c r^2} + \frac{\ddot{\mathbf{\mu}}(t)}{c^2r} \right] + \mathbf{e_{\mathrm{r}}} \mathbf{e_{\mathrm{r}}} \cdot \left[ 3 \frac{\mathbf{\mu}(t)}{r^3} + 3 \frac{\dot{\mathbf{\mu}}(t)}{c r^2} + \frac{\ddot{\mathbf{\mu}}(t)}{c^2r}\right].
    \label{eq0}
\end{equation}
Where $\mathbf{\mu}(t)$ denotes the magnetic moment, $\mathbf{e_{\mathrm{r}}} = \mathbf{r}/r$ and $c$ correspond to the unit vector of the radial direction and the speed of light, respectively. The three components of the rotating dipole field in the Cartesian coordinates have been given in \cite{2000ApJ...537..964C}. Following \cite{2000ApJ...537..964C}, we use the Runge-Kutta integration to follow the magnetic field line in space to obtain the three-dimensional magnetosphere of this pulsar. Detailed calculations for the three-dimensional pulsar magnetosphere based on the Runge-Kutta integration have been investigated in previous works \citep[e.g.,][]{1995ApJ...438..314R,2000ApJ...537..964C,2008ApJ...680.1378H,2019SCPMA..6259505L,2024ApJ...963...65W,2024AN....34540010W}.

\par Under the rotating magnetic dipole approximation, we have the three-dimensional pulsar magnetosphere of PSR B1929$+$10. The left panel of Fig. \ref{pulsar3d}(a) depicts the last closed field lines in the three-dimensional space. Due to the relativistic effects, the magnetic field lines become extremely distorted at high altitudes, yielding the magnetosphere of this pulsar behaves as a structure of a ``peanut''-like. As Fig. \ref{pulsar3d}(a) shows, the black field line becomes extremely elongated within the pulsar magnetosphere. It behaves as neither circular nor elliptical, projecting a look down on the rotation axis. It is a representation of the magnetic field lines that have bunching behavior at the low altitudes due to the relativistic effect. Moreover, high altitude distortion impacts the polar cap shape and affects the primary electrons and the secondary electron-positron $e^{\pm}$ pairs' acceleration. The acceleration electric field (i.e., the parallel electric field $\mathbf{E_{||}}$) is sensitive to the magnetic field line within the pulsar magnetosphere \citep{1975ApJ...196...51R}.

\par In Fig. \ref{pulsar3d}(b), we can also see that the field lines originating from the inside of the polar cap bend sharply at certain points and become strongly distorted at the high-altitude magnetosphere. A high altitude distortion can result in the direction of the emission points of some magnetic field lines that are still parallel to the LOS inside the corotating frame of the magnetosphere. This emission geometry yields that some magnetosphere regions at high altitudes still have radio waves contributing to the radio emission in the observed profile of this pulsar. This indicates that the magnetosphere is still active at extremely high altitudes, resulting in high-altitude magnetospheric radio emissions and a large emission beam responsible for the radio emission in the observed profile of this pulsar. The actual pulsar magnetosphere is still not well understood, but there are many calculations used to study the actual pulsar magnetosphere \citep[e.g.,][]{1999ApJ...511..351C,2006ApJ...648L..51S,2009A&A...496..495K,2010ApJ...715.1282B,2012ApJ...746...60L,2012ApJ...749....2K,2014ApJ...785L..33P,2014ApJ...795L..22C,2018ApJ...855...94P,2018ApJ...857...44K}. For the actual pulsar magnetosphere, some numerical results based on our analysis hardly change, and they still give insight into the pulsar emission physics.

\section{Emission Physics}\label{sec4}
\par The emission physics of radio pulsars is very important for understanding the emission mechanism. However, it is still not clear yet and is an open question for the radio pulsars \citep[e.g.,][]{2017JPhCS.932a2001M,2018PhyU...61..353B}. In this work, numerical calculations based on the RVM geometry and the assumption of the vacuum rotating magnetic dipole field are given to understand the radio emission characteristics of the whole pulse phase emission pulsar. These numerical results shed light on our understanding of the phenomenological radio emission of this pulsar.
\subsection{Polar Cap}
\par The boundary of the polar cap of the pulsar is defined by the footprints of the last closed field lines intersecting with the stellar surface. Fig. \ref{polarCap} shows the polar cap shape of PSR B1929$+$10, implying that the polar cap shape of the rotating magnetosphere approximation becomes complicated and large compared to that of the static magnetic dipole field. The shape and size of the polar cap are highly sensitive to the behavior of the last closed field lines close to the light cylinder. At high altitudes, significant distortion occur in the trajectories of some last closed field lines (Fig. \ref{pulsar3d}), this phenomenon results in the boundary of the polar cap in the rotating dipole approximation being smaller than that of in the static dipole approximation, behaving as a discontinuity (or a notch) in the boundary of the polar cap in the rotating dipole approximation and yielding the intersection between the boundary of the polar cap in the static and rotating approximations. Moreover, compared to a simple circle with a radius of $R_{\mathrm{pc}} = R \sqrt{R \Omega/c}$ for a static aligned rotator of radius $R$, where $\Omega$ and $c$ correspond to the angular frequency and the speed of light, respectively \citep[e.g.,][]{1971ApJ...164..529S,1975ApJ...196...51R,2024ApJ...963...65W}, the polar cap shape of the static approximation becomes more elliptical with the inclination angle $\alpha$ increases. Furthermore, a significant compressed shape is noticeable along the latitudinal direction \citep[e.g.,][]{1990MNRAS.245..514B}, when the inclination angle increases. For the polar cap of PSR B1929$+$10, the effect of the compression along the latitudinal direction also influences its polar cap shape.

\par The polar cap shape of the pulsar influences the primary electrons and the secondary electron-positon $e^{\pm}$ pairs production and acceleration under the RS75-type model. An actual structure of the inner polar gap of the rotating dipole is still not well understood, it must be close to the polar cap shape \citep[e.g.,][]{1975ApJ...196...51R}. Moreover, the acceleration of the charged particle is highly sensitive to the potential drop and the curvature of the field line in a pulsar magnetosphere \citep[e.g.,][]{1993ppm..book.....B,2004ApJ...606L..49Q,2012hpa..book.....L,2016era..book.....C}. The behavior of the field lines within the polar cap can affect the potential drop and alter the acceleration of the electric field. A wide polar cap size is seen in the oblique rotator, and it may easily result in a large potential drop. According to the pulsar theory, a zone in the pulsar magnetosphere can accelerate the charged particles, as long as the electron density, $\rho_e$, of this zone, deviating from the Goldreich-Julian density $\rho_{\mathrm{GJ}}$ \citep[e.g.,][]{1969ApJ...157..869G,1975ApJ...196...51R}. This deviation yields the parallel electric field, $E_{||}$, of this zone does not vanish. As the inclination angle increases, compared to the $\alpha=0$ assumption of the model that is introduced by \cite{1969ApJ...157..869G}, the field lines become strongly distorted at high altitudes. This phenomenon may cause a zone where the electron density deviates from $\rho_{\mathrm{GJ}}$, which contributes to an active magnetosphere at the high altitude and is responsible for the profile longitudes with extremely weak emission for the whole 360$^{\circ}$ of longitude emission pulsar.

\subsection{Sparking Pattern}
\par The pattern of the sparks above the polar cap surface gives insight into the magnetospheric radio emission of the pulsars. Under the assumption that only the field lines within the polar cap (i.e., the open field lines) are responsible for the radio emission of the pulsars, we investigate each of the magnetic field lines inside the polar cap and follow them in space to further understand the radio emission feature of the whole 360$^{\circ}$ of longitude emission pulsar. Then we compute the direction of each emission point. Similar to the calculation introduced in \cite{2024ApJ...963...65W,2024AN....34540010W}, we determine the emission point of each open field line and study whether its direction is parallel to the LOS according to the geometric relation $\zeta = \cos^{-1} (\mathbf{\hat{u}_z}/ \sqrt{\mathbf{\hat{u}_x}^2 + \mathbf{\hat{u}_y}^2 + \mathbf{\hat{u}_z}^2})$. Where $\mathbf{\hat{u}} = \mathbf{\hat{u}_x} + \mathbf{\hat{u}_y} + \mathbf{\hat{u}_z}$ is the unit vector of the actual velocity of the charged particle at the emission point. This determination can give the emission point because the magnetic field lines in the corotating frame of the pulsar magnetosphere have no more than one emission point where the emission direction is parallel to the LOS before they intersect the light cylinder \citep[e.g.][]{1975ApJ...196...51R,2019SCPMA..6259505L,2024ApJ...963...65W}. All emission points can be calculated within the pulsar magnetosphere. These emission points produce radio waves that are detectable by the radio telescope. To investigate the sparks above the polar cap surface, we map the pattern of sparks onto the polar cap surface. After taking the effect of the propagation delays in the pulsar magnetosphere into account, we attain the sparking pattern of this pulsar.

\par In this work, we first consider the effect of the aberration and then compute some of the numerical results. Moreover, it is, at least, worth mentioning that the effect of the curved spacetime near the surface of the neutron star can also influence the sparking pattern when the emission region is close to the pulsar. However, for this pulsar, the contribution of this effect results in the phase shift is about 0$^{\circ}$.14. This shift is small compared to the above major effects \citep[e.g.,][]{1994ApJ...425..767G,1997A&A...322..846K}, and can be neglected. The sparking pattern of the two-pole model is responsible for the whole 360$^{\circ}$ of the longitude emission feature of this pulsar. In this work, under the RVM geometry of the inclination angle $\alpha = 55^{\circ}.62$ and the viewing angle $\zeta = 109^{\circ}.09$, we only give all possibilities of the emission zone in the polar cap surface that can be detected by the telescope, and do not select a special region of the polar cap surface to discuss its contribution to the radio emission feature of this pulsar. After taking the effect of the retardation into account, the sparking pattern on the polar cap surface is shown in Fig. \ref{polarcap}.

\par From Figs. \ref{rvm_final} and \ref{polarcap}, we find that the radio emission of the weak component only comes from a narrow and fixed zone above the polar cap surface. Under this sparking pattern, the huge difference in the observed flux density between the main pulse and the interpulse can be naturally understood. This significant difference is due to the emission zone responsible for the radio emission of the interpulse being extremely narrow on the polar cap surface (shown by red and deep red patches), in comparison to the main pulse. The sparks that contribute to the radio emission of the main pulse can easily be created and accelerated because an extremely wide zone on the polar cap surface can be responsible for the radio wave of the main pulse. Moreover, in this work, we find the two new emission components $C_3$ and $C_4$ with the observed flux density is about 10$^{-4}$ of the magnitude of the main pulse, their corresponding emission zone is also extremely narrow on the polar cap surface (pink and light pink patches). Our analysis supports that only the two-pole model can be responsible for the whole 360$^{\circ}$ of longitude emission pulsar and gives insight into the emission zone above the polar cap surface that is used to understand its emission morphology.
 A wide emission zone on the polar cap surface can result in a large potential drop and then alter preferential discharge. The actual sparking pattern may be extremely complex and is close to the conduction of the pulsar magnetosphere \citep[e.g.,][]{2006ApJ...648L..51S,2015ApJ...801L..19P}.

\subsection{Emission Altitude}
\par The acceleration mechanism of the charged particle is related to the emission altitude. Moreover, the geometrical/field-configuration of the pulsars is also close to the emission altitude \citep[e.g.,][]{1984MNRAS.211...57D,1991ApJ...370..643B,1997A&A...322..846K}. The height of the radio pulsars has been investigated in various methods \citep[e.g.,][]{1978ApJ...222.1006C,1991ApJ...370..643B,1992ApJ...385..282P,1997A&A...322..846K,2012hpa..book.....L}, suggesting the altitude is about an order of 1000\, km for the normal pulsars that character the duty cycle of about $10 \%$. To understand the whole 360$^{\circ}$ of the longitude emission feature of PSR B1929$+$10, we calculate the emission altitude. We estimate the emission altitude from the coordinates of the emission point, correcting for the effect of aberration. The results are shown in Fig. \ref{points}(a), indicating that the strong emission window of this pulsar (i.e., main pulse) originates from an altitude lower than 0.15$R_{\mathrm{LC}}$. This result agrees with the estimation from other methods \citep[e.g.,][]{1978ApJ...222.1006C,1991ApJ...370..643B,1992ApJ...385..282P,1997A&A...322..846K,2012hpa..book.....L}. However, some of the emission windows such as the ``bridge'' emission window and the two new emission windows, originate from an extremely high-altitude magnetosphere where the altitude is up to $R_{\mathrm{LC}}$ and even exceeds $R_{\mathrm{LC}}$ of this pulsar.

\par Due to the field lines becoming strongly distorted, the radio telescope still detects the radio wave contributed by the emission points whose altitudes are larger than 2$R_{\mathrm{LC}}$ (deep red patches in Fig. \ref{points}(a)). These zones in the magnetosphere of this pulsar can still accelerate the charged particles and then emit the radio wave, they are responsible for the observed flux density in the longitude range of -140 to -60$^{\circ}$ (see Figs. \ref{rvm_final}, \ref{polarcap}, and \ref{points}). Our analysis suggests the possibility that some of the zones with extremely high altitudes within the magnetosphere of this pulsar still contribute to the secondary electron-position $e^{\pm}$ pairs creation and acceleration since the density of the primary electrons that are produced in the inner polar gap and their acceleration mechanism in the height of about 2$R_{\mathrm{LC}}$ are a big challenge. These extremely high-altitude magnetosphere zones are responsible for the radio signal of the observed profile longitude with extremely weak emission. These results strongly suggest that the magnetosphere of the whole 360$^{\circ}$ of longitude emission pulsar, PSR B1929$+$10, is still active at an extremely high altitude. An active high-altitude magnetosphere demonstrates that this pulsar has high-altitude magnetospheric radio emissions. Our analysis suggests that the magnetic dipole field is presumably dominant in the radio emission of this pulsar.

\par An active high-altitude magnetosphere significantly influences the extremely weak radio emissions and presents substantial challenges to our understanding of the production and acceleration mechanisms of the charged particles under the RS75-type model. The RS75-type model investigates the physical mechanism of the primary electrons in the inner polar gap with an altitude of about 10$^{2}$\;m \citep{1975ApJ...196...51R}. Two normal radio pulsars, PSRs B0950$+$08 and B1929$+$10, are found to the radio emission detected over all of the pulse period \citep{2022MNRAS.517.5560W}. Moreover, most of the millisecond pulsars (MSPs) also have extremely wide observed profiles \citep[e.g.,][]{2004ApJ...609..354M,2015MNRAS.449.3223D}. Some of the magnetar characterize the $100 \%$ duty cycle in their radio observed pulse profiles \citep[e.g.,][]{2010ApJ...721L..33L}. Our analysis suggests that there are acceleration zones in high altitude magnetosphere that are responsible for the radio emission of the extremely weak emission windows of the pulsars with the observed profile occupying most and all of the pulse period \citep[e.g.,][]{2004ApJ...609..354M,2015MNRAS.449.3223D,2022MNRAS.517.5560W}. The physical mechanism of the outer gap model has been proposed and used to understand the high-energy emission from the pulsars \citep[e.g.,][]{1986ApJ...300..500C,1997ApJ...487..370Z,2000ApJ...537..964C}. Other acceleration zones such as the slot gap \citep[e.g.,][]{1983ApJ...266..215A,2003ApJ...588..430M,2003ApJ...598.1201D,2008ApJ...680.1378H} and the annular gap \citep{2004ApJ...606L..49Q,2012ApJ...748...84D,2024ApJ...963...65W} models are still possible, and they are responsible for the observed profile longitudes with extremely weak radio emission. A new type of acceleration zone in the high-altitude magnetosphere is still possible, this possibility may be evidence of the whole 360$^{\circ}$ of longitude emission pulsar that characterizes a large inclination angle.

\subsection{Curvature Radius}
\par An active high-altitude magnetosphere suggests the radio emission of the strongly distorted field lines. Charged particles are accelerated along the curved magnetic field lines and then emit radio waves under the framework of the magnetic dipole field \citep[e.g.,][]{1971ApJ...164..529S,1975ApJ...196...51R,1979ApJ...231..854A}. The curvature of the field lines directly reveals important information about the emission morphologies of pulsars. Compared with the static dipole field, the magnetic field line of the oblique rotator must become complicated. The magnetic field lines bend sharply in certain regions of the magnetosphere under the assumption of the rotating dipole approximation (see Figs. \ref{pulsar3d}). Under the pulsar magnetosphere of the rotating dipole approximation, we present the numerical result of the curvature radius of the pulsar. We estimate the curvature radius of the emission point based on its definition.
\begin{equation}
    \rho = \frac{ds}{d\theta}.
    \label{eq2}
\end{equation}
Here $ds$ denotes the arc length corresponding to the angle of $d\theta$. Our results are shown in Fig. \ref{points}(b), one can see that the curvature radius in the regions near the boundary of the polar cap is usually low. A low curvature radius indicates a strongly distorted magnetic field line at that emission point and a significant effect of the rotation of the neutron star on the field lines. However, the curvature radius of the regions becomes large while they are close to the magnetic pole, which means a bend slightly in their corresponding magnetic field lines. The curvature radius of some regions becomes extremely large and is up to 2$R_{\mathrm{LC}}$ (the crimson patches). This phenomenon is because the magnetic field lines have little distortion at the high-altitude magnetosphere as the inclination angle increases at their corresponding emission points.

\par The curvature radius of the magnetic field line not only affects the acceleration trajectories of the primary electrons but also influences the distribution of the electric and magnetic fields in the pulsar magnetosphere \citep{1975ApJ...196...51R,1993ppm..book.....B,2022ARA&A..60..495P}. The different curvature radii at different regions of the polar cap surface may contribute to the radio emission at different frequencies \citep[e.g.,][]{1978ApJ...222.1006C,2016era..book.....C}. Our analysis suggests that the strongly distorted field lines are responsible for the high-altitude magnetospheric radio emission when the inclination angle becomes large. Moreover, the curvature radius of the emission point is not directly related to the emission altitude. From Figs. \ref{points}(a) and (b) we can see that some of the emission regions where the magnetic field lines behave as weakly distorted corresponding to very large values of the curvature radius but their emission altitudes are not high enough and not near the magnetic pole (see the crimson patches in the Fig. \ref{points}(b)). Our result suggests that the rotating dipole magnetosphere yields the relativistic effects that have a significant effect on the magnetic field lines, resulting in a large radio emission beam at high-altitude magnetosphere regions. These results are responsible for the radio emission of the pulsars whose observed profiles occupy most or all of the pulse period when the inclination angle increases.

\section{Discussion}\label{sec5}
\par The emission mechanism of radio pulsars is still a matter of debate. The intrinsic radio emission from the pulsars and their corresponding polarization property would give insight into understanding the emission mechanism. Highly polarized emission seen in the whole 360$^{\circ}$ of longitude emission pulsar, PSR B1929$+$10. This is different from that of the whole pulse phase emission pulsar, PSR B0950$+$08, characterizing a low linear polarization factor in the main pulse \citep[][]{2022MNRAS.517.5560W,2024ApJ...963...65W}. Previous observations find that due to a small emission beam, the normal pulsars' observed pulse profiles are usually narrow (about $10 \%$ duty cycle) compared to an extremely wide observed profile in millisecond pulsars (MSPs) and a few magnetars in the radio, the X- or Gamma-rays profiles \citep[e.g.,][]{1997A&A...322..846K,2004ApJ...609..354M,2010ApJ...721L..33L,2015MNRAS.449.3223D}. Our result of detecting the whole pulse phase emission characteristics of PSR B1929$+$10 not only enhances the sample of the whole pulse phase emission pulsars that observationally constrain on the pulsars' models but also gives a good chance to hint at the connection between the pulsars that have near or a $100 \%$ duty cycle regardless of the evolution of these objects. Because these objects have similar emission properties in their observed profiles at different energies. We expect that the sample of the $100\%$ duty cycle in the radio emission is as much as in the X-ray or Gamma-ray emission, as the ability of the observation of the telescope increases.

\par Expect for the extremely wide observed pulse profiles, the MSPs and a few magnetars usually characterize highly polarized emission property \citep[e.g.,][]{2006MNRAS.368.1856J,2007MNRAS.377..107K,2010ApJ...721L..33L,2015MNRAS.449.3223D}. The radio emission characteristics of PSR B1929$+$10 are similar to those of the magnetar, PSR J1622$-$4950 \citep{2010ApJ...721L..33L} no matter of the observed pulse profile and the polarization property. Both objects characterize highly polarized emission, and the observed pulse profile occupies almost or the full pulse period. A large pulse duty cycle of PSR J1622$-$4950 is interpreted as a non-dipolar magnetic field structure in its magnetosphere. However, the magnetic dipole field structure is usually responsible for the emission property of the normal radio pulsars. Similar radio emission properties seen in the two objects may indicate that the contributions of the magnetic dipole and non-dipole field structures to the magnetospheric emission are related to the magnetic field strength since the magnetar characterizes a high magnetic field and is powered by the energy stored in its large magnetic field, typically $\gtrsim 10^{14}$\;G \citep[e.g.,][]{1992ApJ...392L...9D,2010ApJ...721L..33L}. We argue that these connections between the MSPs, the magnetars, and the normal pulsars would be more ordinary with the detection ability increasing in the future.

\par The RVM geometry is based on the variation of the orientation of the magnetic field lines of the pure dipolar magnetic field with respect to the $\Vec{\Omega}-\mu$ plane \citep{1969ApL.....3..225R}. It predicts a monotonic rotation of the position angle across the pulsed emission phase and behaves as a ``S''-shape. This model successfully predicts the swing of the observed PPAs of the Vela pulsar (PSR B0833$-$45). Moreover, a large number of polarization observations of the pulsars that characterize low-duty cycle ($\sim 10 \%$) suggest that the RVM geometry is successfully responsible for their emission geometries. However, with the sample of the pulsars that characterize unusually wide observed pulse profiles increases, particularly in the MSPs, clear deviations are seen in the observed PPA variations over pulsed emission window and the framework of the RVM geometry \citep[e.g.,][]{2004ApJ...609..354M,2015MNRAS.449.3223D,2023MNRAS.520.4801J,2023RAA....23j4002W}. For the pulsars that have a large separation between their main pulse and interpulse, previous polarization measurements suggest that their observed PPAs variations in the pulse longitude usually need to perform twice independently RVM fits for the main pulse and the interpulse, respectively \citep[e.g.,][]{2001ApJ...553..341E,2006MNRAS.366..945W,2007MNRAS.377..107K,2012hpa..book.....L}. For the whole 360$^{\circ}$ of longitude emission pulsar, PSR B1929$+$10, we measure the observed PPAs of this pulsar throughout the full rotation period. The observed PPA variations in the full pulse longitude give a good opportunity to test the RVM geometry. In this work, we fit the observed PPA variations by the RVM in the full longitude (see Fig. \ref{rvm_final}). Our result indicates that the swing of the observed PPA over the full longitude can be described by the framework of the RVM geometry well after taking the orthogonal polarization modes in the pulse longitude range (-27$^{\circ}$, 6$^{\circ}$) that yields a 90$^{\circ}$ jump in the observed PPA into account. Our analysis indicates that the RVM framework can still account for the emission geometry of this pulsar, which has full pulse longitude emission characteristics. Our result observationally supports the framework of the rotating vector model \citep{1969ApL.....3..225R} for the pulsars that characterize extremely wide observed pulse profiles and have highly linear polarized profiles.

\par It is worth noticing, as mentioned above, that the whole pulse phase emission pulsar PSR B1929$+$10 is characterized by about $100 \%$ linear polarization in most pulse longitudes (see Fig. \ref{rvm_final}). The phenomenon suggests that the conversion of the linear into circular polarizations in the magnetosphere of this pulsar is not effective, and then yields a well-defined RVM geometry. For the pulsars that have the complicated observed profiles, their weak emission windows, such as the pre- and post-cursors components and the ``bridge'' that generally characterize a near $100 \%$ fractional linear polarization \citep[e.g.,][]{2004ApJ...609..354M,2015MNRAS.449.3223D}. The phenomenon that highly linear polarized seen in these emission windows is higher than that of the strong emission windows (i.e., main pulse) is commonly seen in the polarization emission behavior of the pulsars, and its origin is still unclear. The physical mechanism of this phenomenon may be hinted at by the analysis of the emission physics of this pulsar, which shows that high-altitude magnetosphere regions are responsible for the origin of the highly polarized emission property.

\par The emission location of the pulsars significantly influences the understanding of the emission mechanism. For the pulsars that behave as a narrow observed pulse profile ($\sim 10 \%$ duty cycle), a large number of estimations of the emission location give that the primary electrons contribute to the radio emission of them in the low-altitude magnetosphere region \citep[e.g.,][]{1975ApJ...196...51R,1977puls.book.....M,1978ApJ...222.1006C,1991ApJ...370..643B,1992ApJ...385..282P,1997A&A...322..846K,2012hpa..book.....L}. For high-energy pulsars usually exhibit extremely wide observed profiles that cover most of the rotation phase, with some occupying the full rotation phase. The emission mechanisms, such as the curvature, synchrotron, and inverse Compton radiation, relate to the physical processes in the pulsar magnetosphere of primary electrons and secondary electron-position $e^{\pm}$ pairs. These mechanisms are responsible for the emission features of these objects in the high-altitude magnetosphere region \citep[e.g.,][]{1983ApJ...266..215A,1986ApJ...300..500C,2007ChJAA...7..496Q,2008ApJ...680.1378H}.

\par As investigated in the previous works \citep[e.g.,][]{1983ApJ...266..215A,1986ApJ...300..500C,2007ChJAA...7..496Q,2008ApJ...680.1378H}, a high-altitude emission yields a large emission beam that results in a much larger area sky. The different types of accelerators, such as the outer, the slot, and the annular gap models, contribute to the primary electrons and secondary electron-position $e^{\pm}$ pairs. This emission beam geometry is responsible for the $100 \%$ duty cycle in the high-energy observed profile. For the low-altitude magnetosphere emission, an aligned rotator can also naturally explain the whole 360$^{\circ}$ of longitude emission characteristics. This emission geometry is evidence of the radio emission characteristics of PSR B0826$-$34 extending through the whole pulse phase when it switches into the ``strong emission state'' \citep{2005MNRAS.356...59E}. However, the polarization measurement of this pulsar demonstrates that it is an oblique rotator that has the inclination angle $\alpha$ of about 56$^{\circ}$.

\par Similar to the previous investigations \citep[e.g.,][]{1995ApJ...438..314R,2000ApJ...537..964C,2008ApJ...680.1378H,2019SCPMA..6259505L}, we utilize the retarded vacuum dipole solution to attain the magnetosphere structure of this pulsar. Under the rotating dipole approximation, we find that the weak emission windows come from extremely high-altitude magnetosphere regions (see Figs. \ref{rvm_final}, \ref{polarcap}, and \ref{points}), indicating that the high-altitude magnetosphere of this normal radio pulsar is still active. As discussed above, a high-altitude emission yields a large emission beam. This deviates from the scale of the emission beam of the normal radio pulsar with $\propto 1/ \sqrt{P}$. Here $P$ is the period of pulsar. Our quantitative results based on the numerical calculation would shed light on the understanding of the emission mechanism of the pulsars that characterize most or all of the pulse period emission property regardless of their emission energies, i.e., the radio, X-ray, and Gamma-ray. A unified pulsar's model for the class of pulsars that have extremely wide observed profiles is surely a requirement since the sample of these objects, particularly in the MSPs, has significantly increased \citep[e.g.,][]{2004ApJ...609..354M,2010ApJ...721L..33L,2015MNRAS.449.3223D,2022MNRAS.517.5560W,2023MNRAS.520.4801J,2023RAA....23j4002W}.

\par Radio waves are emitted in multi-narrow bands for several pulses, as shown in Fig. \ref{drift}.
These narrowband structures may be caused by a quasi-periodically distributed bunches, which are triggered by a quasi-periodic spark in the inner polar gap \citep[e.g.,][]{2023ApJ...956...67Y,2024A&A...685A..87W}.
The extracted electrons in the gap can screen the gap once the number density is high enough, and the sparking process stops.
When the plasma moves along the magnetic field and leaves the gap region, the gap with almost no particles inside is approximately formed, and the vacuum electric field is no longer screened. The sparking process works again. Such a whole process is quasi-periodic.
Based on the numerical simulation results given by \cite{2015ApJ...810..144T}, the periodicity of the whole process is $\sim3h_{\rm gap}/c$, where $h_{\rm gap}$ is the height of the gap.

\par Two radio pulsars, B0950$+$08 and B1929$+$10, have the whole pulse phase emission characteristics. The polarization measurements obtained with FAST show that both objects have large inclination angles. Specifically, the angle is $\alpha=100^{\circ}.5$ for B0950$+$08 \citep{2024ApJ...963...65W} and $\alpha=55^{\circ}.62$ for B1929$+$10. This indicates that their emission geometry are not aligned rotators, where the inclination angle would be close to $0^{\circ}$, nor orthogonal rotators, where the inclination angle would approach $90^{\circ}$. Phenomenological pulsar models based on the magnetic dipole field include a low-altitude accelerator, such as the inner polar gap, which is responsible for the primary charged particles and interprets their radio emission features. The radio emission behaviors of these two objects characterize the full pulse period, which enhances our understanding of pulsars' emission with the pulse property in the radio wave. According to the investigation of emission geometry and pulsar magnetosphere for PSRs B0950$+$08 and B1929$+$10, the strong pulse component, i.e., the main pulse, is emitted from the low-altitude magnetosphere region. A low-altitude magnetospheric radio emission can be understood under the pulsar models. However, the weak emission components come from the high-altitude magnetosphere, which is a big challenge for the pulsar models that are based on the inner polar gap acceleration mechanism to understand this radio emission. This is due to the significant challenges of the electric field strength and the acceleration mechanisms of charged particles in the high-altitude magnetosphere.

\par The emission signals of the high-energy pulsars, such as the X-ray and some of the Gamma-ray pulsars, are usually detected over their entire rotation periods. Similar emission features are observed in the pulsars at different energies, which may hint at similar physical processes in their magnetospheres. In addition, the detection of the whole pulse phase emission feature from these two bright pulsars, PSRs B0950$+$08 and B1929$+$10, encourages us to conduct a campaign to long-term monitor other bright pulsars by FAST to identify whether they have radio emission features similar to B0950$+$08 and B1929$+$10. Further research and monitoring of pulsars with unusually wide observed profiles are valuable, as they can observationally constrain pulsar model and shed light on the underlying connections of these objects in their physical processes, even though they emit at different energies.

\section{Conclusion}\label{sec6}
\par In this work, we detect two new emission components in the observed profile of PSR B1929$+$10. The observed flux densities of these two components are extremely weak and are about 10$^{-4}$ of the magnitude of the peak radio emission (see Fig. \ref{f1}). Our result is strong evidence that PSR B1929$+$10 is a whole 360$^{\circ}$ of longitude emission pulsar. Moreover, its average observed profile detects an extremely complex pulse component with at least 15 emission components. We investigate the radio emission behaviors of these 15 pulse components by using a power-law function to fit them and find that the strong pulse components, i.e., main pulse and interpulse, have a flat spectral index of less than 1.7. Interestingly, the weak pulse components character a steep spectral index larger than 2.3 (see Table \ref{t1}). Multi-component emission structures (i.e., 10 or even 12 components) are seen in a few MSP profiles \citep[e.g.,][]{1998ApJ...501..270K,2004ApJ...609..354M} but are extremely rare in the observed profile of the normal pulsars. The different spectral indices may hint at different physical mechanisms. Five modes of subpulse modulation have been identified based on the analysis of the LRFS, with Modes 2, 4, and 5 being first confirmed in this observation. The detection that the mode of subpulse modulation is lower than the pulse components indicates that some of the pulse components have merged in the physical process of subpulse modulation.

\par Narrowband emission feature is most seen in FRB's emission and a sudden jump in the observed PPAs recently detected in FRBs \citep[e.g.,][]{2024ApJ...972L..20N}. These two emission features are also detected in the single pulse of this pulsar. This suggests that the underlying physical mechanism of some FRBs is similar to that of pulsars. The radio emission of these two objects originates from the magnetosphere. Understanding the physical mechanism of frequent jumps in the observed PPAs is still a big challenge. In addition, the jump in the observed PPAs fails to be described by the RVM geometry \citep{1969ApL.....3..225R} and must be close to the complex physical processes of the radio wave propagation in the magnetosphere  \citep[e.g.,][]{1986ApJ...302..120A,2010MNRAS.403..569W}.

\par Similar to studying the whole 360$^{\circ}$ of longitude emission pulsar, PSR B0950$+$08 \citep{2022MNRAS.517.5560W,2024ApJ...963...65W,2024AN....34540010W}, we investigate the emission geometry of PSR B1929$+$10 by fitting the observed PPAs of the averaged pulse by the RVM. We obtain two sets of $\alpha,\zeta$ from the different fit routines and discuss them. We find the solution of the set of ${\alpha = 55^{\circ}.62, \zeta = 109^{\circ}.09}$ is a better one as its emission geometry can be responsible for the huge difference observed flux density at different pulse longitudes of this pulsar. This RVM solution shows that the whole 360$^{\circ}$ of longitude emission pulsar, PSR B1929$+$10, is an oblique rotator instead of a nearly aligned rotator. \citep[e.g.,][]{1988MNRAS.234..477L,1990ApJ...361L..57P,1991ApJ...370..643B,1997JApA...18...91R,2001ApJ...553..341E}. This emission geometry can naturally understand the huge difference in the observed flux density between the main pulse and the interpulse. A large viewing angle results in a large emission beam.

\par Under the assumption of the retarded vacuum magnetic dipole field \citep[e.g.,][]{1995ApJ...438..314R,2000ApJ...537..964C,2008ApJ...680.1378H,2019SCPMA..6259505L}, we investigate its magnetosphere geometry using the RVM solution of $\alpha = 55^{\circ}.62, \zeta = 109^{\circ}.09$ in the rotating approximation. In Fig. \ref{pulsar3d}(a), one can see that the last closed field line becomes extremely complicated instead of a simple sphere in space. Some of the last closed field lines exhibit elongation, which causes the magnetosphere to behave like a ``peanut-like'' structure. This irregular structure due to the magnetic field lines becomes strongly distorted at high altitudes and the effect of the magnetic field line sweepback when the inclination angle becomes large, resulting in some regions on the polar cap surface where the field lines become extremely elongated (see the left of Fig. \ref{pulsar3d}(a)). In the right of Fig. \ref{pulsar3d}(a), it is also visible in the way of projection looking down. These phenomena yield a discontinuity (or a notch) in the boundary of the polar cap surface (see Fig. \ref{polarcap}). Moreover, the analysis of the sparking pattern indicates that the weak emission windows, as well as two new components, only originate from the narrow zone in the polar cap surface, and their corresponding emission altitude is extremely high. In contrast, the radio emission of the strong pulse phase (i.e., main pulse) comes from a wide zone with a low-altitude magnetosphere region. These results agree with the huge difference in the observed flux density between the main pulse and other pulse longitudes for the radio emission feature of the whole 360$^{\circ}$ of longitude emission pulsar. This is because the potential drop is close to the width of the emission zone of the polar cap surface \citep[e.g.,][]{1975ApJ...196...51R,2004ApJ...606L..49Q,2024ApJ...963...65W}.

\par Our analysis suggests that the high-altitude magnetosphere of the whole 360$^{\circ}$ of longitude emission pulsar, PSR B1929$+$10, is still active. An active high-altitude magnetosphere is responsible for the radio emission of weak radio emission windows and their corresponding highly polarized emission of this pulsar. These results suggest that this pulsar has a high-altitude magnetospheric radio emission. The magnetic field lines are extremely distorted, resulting in a large emission beam. Our analysis indicates that high emission altitudes do not directly relate to a large curvature radius in certain magnetosphere regions. In Fig. \ref{points} we can see that some of the emission pulse phases come from the locations with an altitude of about 0.6$R_{\mathrm{LC}}$ but have an extremely large curvature radius (crimson patches in Fig. \ref{points}(b)).

\begin{acknowledgements}
      This work is supported by the Strategic Priority Research Program of the Chinese Academy of Sciences (Nos. XDA0350501, XDB0550300), the National Natural Science Foundation of China (12133003), the Guangxi Talent Program (Highland of Innovation Talents), the National Key R$\&$D Program of China (Grant No. 2024YFA1611700), the National SKA Program of China (2020SKA0120100), the Guizhou Province High-level Talent Program (No.QKHPTRC-GCC[2022]003-1), and the Guizhou Provincial Science and Technology Projects (Nos. QKHFQ[2023]003, QKHFQ[2024]001, QKHPTRC-ZDSYS[2023]003). This work made use of the data from the Five-hundred-meter Aperture Spherical radio Telescope (FAST), operated by the National Astronomical Observatories, Chinese Academy of Sciences.
\end{acknowledgements}

\bibliography{aa}{}
\bibliographystyle{aa}

\end{document}